%% file: diss_revision.tex
\documentclass[review,nonatbib]{elsarticle}
\usepackage[a4paper, margin=1in]{geometry}

\usepackage[nolist]{acronym} 			
\usepackage{amsmath}
\usepackage{amssymb}
\usepackage{crimson}
\usepackage{comment}
\usepackage{graphicx} 
\usepackage{url}
\usepackage{subcaption}
\usepackage{floatrow}
\usepackage{float}
\usepackage{booktabs}
\usepackage{longtable}

\usepackage{hyperref}
\hypersetup{colorlinks=true}

\usepackage{multirow}

\usepackage{caption}
\makeatletter
\let\c@author\relax
\makeatother

\usepackage[
    backend=biber,
    isbn=false,
    url=false,
    sorting=none,
    citestyle=numeric-comp,
    doi=true,
    natbib=true
    ]{biblatex}
\journal{arXiv}

\begin{acronym}
 \acro{LDES}{long-duration electricity storage}
 \acro{LDTS}{long-duration thermal storage}
 \acro{TWh}{terawatt-hours}
 \acro{DH}{district heating}
 \acro{COP}{coefficient of performance}
 \acro{MCRL}{maximum cumulative residual load}
 \acro{VRE}{variable renewable energy}
\end{acronym}

\begin{document}

\begin{abstract}
Hydrogen-based \acf{LDES} is a key component of renewable energy systems to deal with seasonality and prolonged periods of low wind and solar energy availability. In this paper, we investigate how electrified heating with heat pumps impacts \ac{LDES} requirements in a fully renewable European energy system, and which role thermal storage can play. Using a large weather dataset of 78 weather years, we find that electrified heating significantly increases \ac{LDES} needs, as optimal average energy capacities more than quadruple across all weather years compared to a scenario without electrified heating. We attribute 75\% of this increase to a leverage effect, as additional electric load amplifies storage needs during times of low renewable availability. The remaining 25\% are the result of a compound effect, where exceptional cold spells coincide with  periods of renewable scarcity. Furthermore, heat pumps increase the variance in optimal storage capacities between weather years substantially because of demand-side weather variability. Long-duration thermal storage attached to district heating networks can reduce \ac{LDES} needs by on average~36\%. To support and safeguard wide-spread heating electrification, policymakers should expedite the creation of adequate regulatory frameworks for both long-duration storage types to de-risk investments in light of high weather variability.
\end{abstract}


\begin{frontmatter}

\title{A mix of long-duration hydrogen and thermal storage enables large-scale electrified heating in a renewable European energy system}



\author[A1,A2]{Felix Schmidt\corref{FS}}
\ead{fschmidt@diw.de}

\author[A1,A3]{Alexander Roth}
\ead{alexander.roth@kuleuven.be}

\author[A1]{Wolf-Peter Schill}
\ead{wschill@diw.de}

\cortext[FS]{Corresponding author}
\address[A1]{DIW Berlin, Department of Energy, Transportation, Environment, Mohrenstra{\ss}e 58, 10117 Berlin, Germany}
\address[A2]{TU Berlin}
\address[A3]{Divison of Applied Mechanics and Energy Conversion, Katholieke Universiteit Leuven, Celestijnenlaan 300, 3001 Leuven, Belgium}

\end{frontmatter}

\section{Introduction}

The global effort to mitigate climate change relies crucially on the decarbonization of the energy system \cite{intergovernmental_panel_on_climate_change_ipcc_technical_2023,international_energy_agency_net_2024}. For many countries, especially in Europe, studies suggest that a power system largely based on variable wind and solar power is feasible and cost-effective to decarbonize the energy system \cite{brown_response_2018,hansen_status_2019,luderer_impact_2022,breyer_history_2022}. 
Planning weather-resilient \ac{VRE} systems requires considering weather variations over long timescales and extreme events \cite{pfenninger_dealing_2017,van_der_wiel_influence_2019,bloomfield_quantifying_2016,craig_overcoming_2022}. A growing strand of the literature analyzes the impact of weather variability on optimal renewable system designs \cite{gotske_designing_2024,ruggles_planning_2024,grochowicz_intersecting_2023,goke_stabilized_2024}, investigates periods of extremely low renewable generation, also referred to as \textit{Dunkelflaute} \cite{kittel_measuring_2024,kittel_coping_2025}, determines flexibility needs to deal with variability and extreme events \cite{lund_review_2015,alizadeh_flexibility_2016,brown_synergies_2018,dowling_role_2020,ruhnau_storage_2022}, and seeks to overcome the computational challenges associated with including decades of weather data in capacity expansion models \cite{goke_stabilized_2024,jacob_future_2023}.

Energy storage technologies are central for bridging periods with insufficient supply of solar or wind energy \cite{schill_electricity_2020,levin_energy_2023}. To provide energy during prolonged Dunkelflaute events, \acf{LDES} has emerged as a key enabler \cite{levin_energy_2023,dowling_role_2020,dowling_long-duration_2021,staadecker_value_2024,jenkins_long-duration_2021}. While a range of \ac{LDES} technologies is potentially available, the key metric for their competitiveness is the cost of energy capacity \cite{sepulveda_design_2021}. Despite geographic restrictions that may put a strain on transmission infrastructure \cite{brown_ultra-long-duration_2023}, the debate has been centered around hydrogen cavern storage \cite{dowling_long-duration_2021}. Model results on optimal \ac{LDES} capacities vary substantially on the selected weather data \cite{kittel_coping_2025,ruhnau_storage_2022}, geographic scope, considered sectors, and other model specifications. 

The direct or indirect electrification of sectors such as transport, industry, and heating is crucial for the complete decarbonization of economies around the world \cite{international_energy_agency_net_2024}. Yet, most studies at the nexus of \ac{LDES} and weather variability focus only on the power sector and neglect the impact of weather variations on the demand side of a sector-coupled renewable system \cite{gotske_designing_2024}. Accounting for 16\% of total greenhouse gas emissions in the European Union in 2024 \cite{rozsai_m_jrc-idees-2021_2024}, the heating sector takes center stage in the continent's decarbonization efforts \cite{commission_renewable_2023}. Previous studies have shown that serving a substantial part of residential and commercial heat demand with heat pumps can have challenging consequences for the power system \cite{bloess_power--heat_2018,zeyen_mitigating_2021,roth_power_2023,roth_power_2024}. Furthermore, there is a substantial degree of interannual variation in temperatures, and correspondingly in heat demand and temperature-dependent heat pump efficiencies. There is evidence that cold spells can coincide with times of renewable supply shortages \cite{van_der_wiel_influence_2019,mockert_meteorological_2023} exacerbating system stress.

\Ac{LDTS} is a potential alternative to \ac{LDES} to deal with heating-related system imbalances. Thermal energy storage has been shown to improve the utilization of renewables and to reduce heat prices by smoothing heat demand peaks in Danish case studies \cite{sifnaios_impact_2023,christensen_role_2024}. Another Danish study finds it to be a no-regret option for flexibility in district energy systems with electrolyzers \cite{ostergaard_optimal_2023}. Short-duration thermal storage can complement batteries \cite{jacob_future_2023,roth_power_2023}. Zeyen et al.~conclude that thermal storage can mitigate heating demand peaks in a sector-coupled European energy system, especially in combination with building renovations \cite{zeyen_mitigating_2021}. A Norwegian case study discusses alternative configurations of \ac{LDTS} in local district heating systems with renewables \cite{kauko_flexibility_2024}. A direct comparison of hydrogen and thermal storage for the Swedish system concludes that hydrogen storage is better suited to integrate wind power, but less effective in reducing fuel consumption \cite{sundarrajan_harnessing_2025}. In a large sector-coupled European system model for the weather year 2015, Victoria et al.~find that electrifying the heating sector comes with large \ac{LDTS} capacities, but these do not replace hydrogen storage capacities \cite{victoria_role_2019}. In conclusion, the the previous literature on heat electrification and \ac{LDTS} is largely focused on local energy systems and usually considers single or few weather years only.

We contribute to the literature with an integrated analysis of long-duration electricity and thermal storage needs in a fully renewable European energy system with large-scale electrified heating. 
We do so at an extensive geographic and temporal scope, using a sector-coupled capacity expansion model spanning 28 European countries at an hourly resolution and 
using a comprehensive dataset of 78~weather years. The latter features weather-dependent time series of wind and solar capacity factors, hydro inflows, heat demand and heat pump efficiencies. 
We exploit the variation between weather years to isolate the drivers of increased \ac{LDES} needs, differentiating between a leverage and a compound effect. We further quantify the potential role of \ac{LDTS} in providing long-duration flexibility and discuss its potential to substitute or complement \ac{LDES}, while most of the existing literature has focused on short-duration thermal energy storage. Based on our findings, we discuss implications for potential storage investors and policymakers.

\section{Results}

In the following, we focus primarily on the link between the electrification of residential and commercial heat and \ac{LDES} needs. The Supplemental Information Section \ref{sec:additional_output} provides several additional results. We summarize all considered scenarios in Table~\ref{tab:scenarios} and provide additional details in Section~\ref{sec:scenario_design}.

\begin{table}[H]
    \centering
    \caption{\textit{Scenarios}}
    \include{tables/scenario_table}
    \label{tab:scenarios}
\end{table}

\subsection{Electrified heat demand leads to significantly higher long-duration electricity storage requirements}

\begin{figure}[H]
    \centering
    \includegraphics[width=\linewidth]{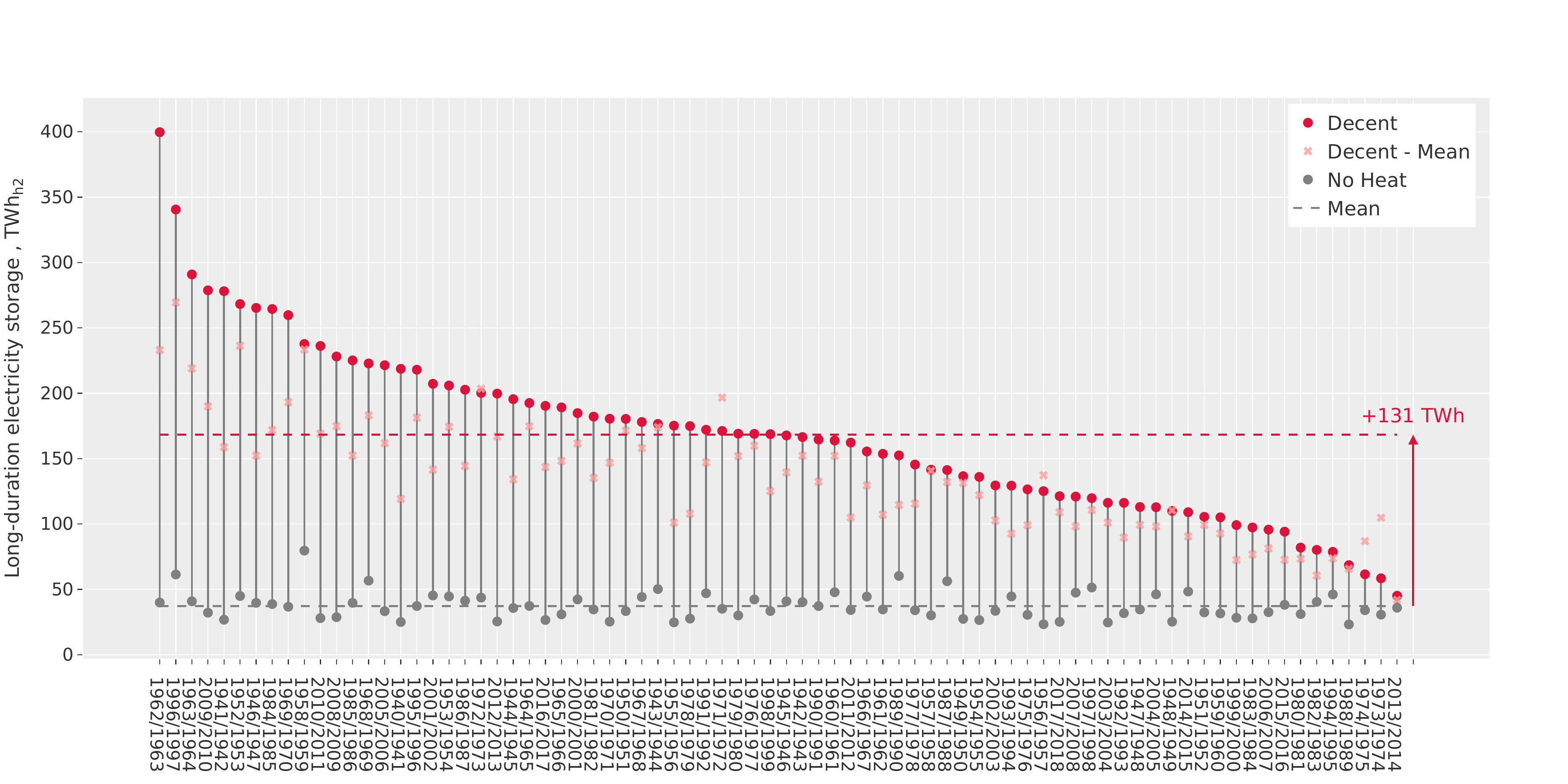}
    \caption{\textit{Long-duration electricity storage requirements in scenarios with different electrified heat demand levels} \\ \scriptsize{Optimal long-duration electricity storage capacities by weather year. Gray dots represent optimal capacities in the \textit{No Heat} scenario. Dark-red dots represent optimal capacities in the \textit{Decent} scenario. Light-red crosses represent optimal capacities in the \textsc{Decent - Mean} scenario, in which weather-year specific heat demand is replaced by long-run mean heat demand.}}
    \label{fig:add_lds}
\end{figure}

To identify how much additional \ac{LDES} energy capacity is needed due to the electrification of heat, we define two sets of scenario runs. In the first set (\textit{No Heat}), the model jointly optimizes capacity and dispatch decisions in a fully renewable system without any electrified heat. The resulting aggregate \ac{LDES} energy capacities for all 28~countries vary by weather year (gray dots in Figure \ref{fig:add_lds}). Optimal aggregate capacities across Europe range between 23~\ac{TWh} in 1988/89 to 80~\ac{TWh} in 1958/59. The average \ac{LDES} capacity across all weather years is 37~\ac{TWh}.

In the second set of runs (\textit{Decent}), we assume that 80\% (see Section \ref{sec:scenario_design}) of all residential and commercial heat demand (space and hot water) is electrified, using decentralized heat pumps with small buffer thermal storage. This increases the mean aggregate \ac{LDES} requirement to 168~TWh, corresponding to an increase of 131~TWh (273\%) on average (represented by the dark-red dots in Figure \ref{fig:add_lds}). In 1962/63, the year with the highest \ac{LDES} requirement in the set of runs with electrified heat, the \ac{LDES} capacity increases tenfold from 40~\ac{TWh} to 400~\ac{TWh}.

In Figure \ref{fig:add_lds}, the weather years are sorted by \ac{LDES} requirement in the \textit{Decent} scenario. Notably, the \textit{No Heat} scenario does not reflect the same order. Weather years that have high \ac{LDES} requirements in the scenario without electrified heat do not necessarily rank high when considering electrified heat (cf. 1989/90) and \textit{vice versa} (cf. 1941/42). Additionally, the distribution of optimal \ac{LDES} capacities widens substantially. In the \textit{No Heat} scenario, the standard deviation of total \ac{LDES} energy capacities across all weather years is 10~TWh. In the scenario \textit{Decent}, the standard deviation increases to 66~TWh.

\subsection{Heat demand drives storage needs through `leverage' and `compound' effects} \label{sec:lev_comp}

In a third set of scenario runs (\textit{Decent - Mean}), we replace the time series of heat demand and heat pump coefficients of performance (COP) in a given weather year by their long-run hourly averages across all weather years in the sample. These curves (see Figure \ref{fig:input_heat}) isolate the inherent seasonality of electricity demand for heating and abstract from cold spells or unusually warm periods. The results of these model runs (light red crosses in Figure \ref{fig:add_lds}) highlight the drivers of increasing \ac{LDES} requirements. Heat demand is seasonal, and electrified heating naturally adds most load in winter months. As extreme system states of \ac{VRE} scarcity mostly occur during winter months \cite{kittel_quantifying_2024,kittel_coping_2025,gotske_designing_2024}, the additional load amplifies the energy deficit, triggering additional \ac{LDES} requirements by functioning as a lever. Across the sample of weather years, this \textit{leverage effect} accounts for 98~TWh or 75\% of the total \ac{LDES} increase on average.

For some years, the \textit{Decent - Mean} scenario results in even higher \ac{LDES} requirements (e.g.,~1973/74) for which unusually mild conditions offset some of this leverage effect. Yet, in other years, the year-specific heat demand leads to far higher \ac{LDES} capacities, as \ac{VRE} scarcities coincide with cold spells in a \textit{compound effect}. While this effect only explains 25\% on average, cold spells have a considerable impact in some of the most extreme weather years.

\begin{figure}[H]
    \centering
    \caption{\textit{\ac{LDES} trajectories with and without electrified heat in selected weather years} \\ \scriptsize{ 
        LDES trajectories shown against daily deviations in heat demand from long-run mean (bright red - mid) and 24-hour rolling renewable availability (green - low). The gray area represents the storage trajectory in \textit{No Heat}. The darker red area shows the additional storage level in \textit{Decent - Mean}, and the lighter red addition represents the incremental storage level from the specific weather year's heat demand (\textit{Decent})}.
    }
    \includegraphics[width=1\linewidth]{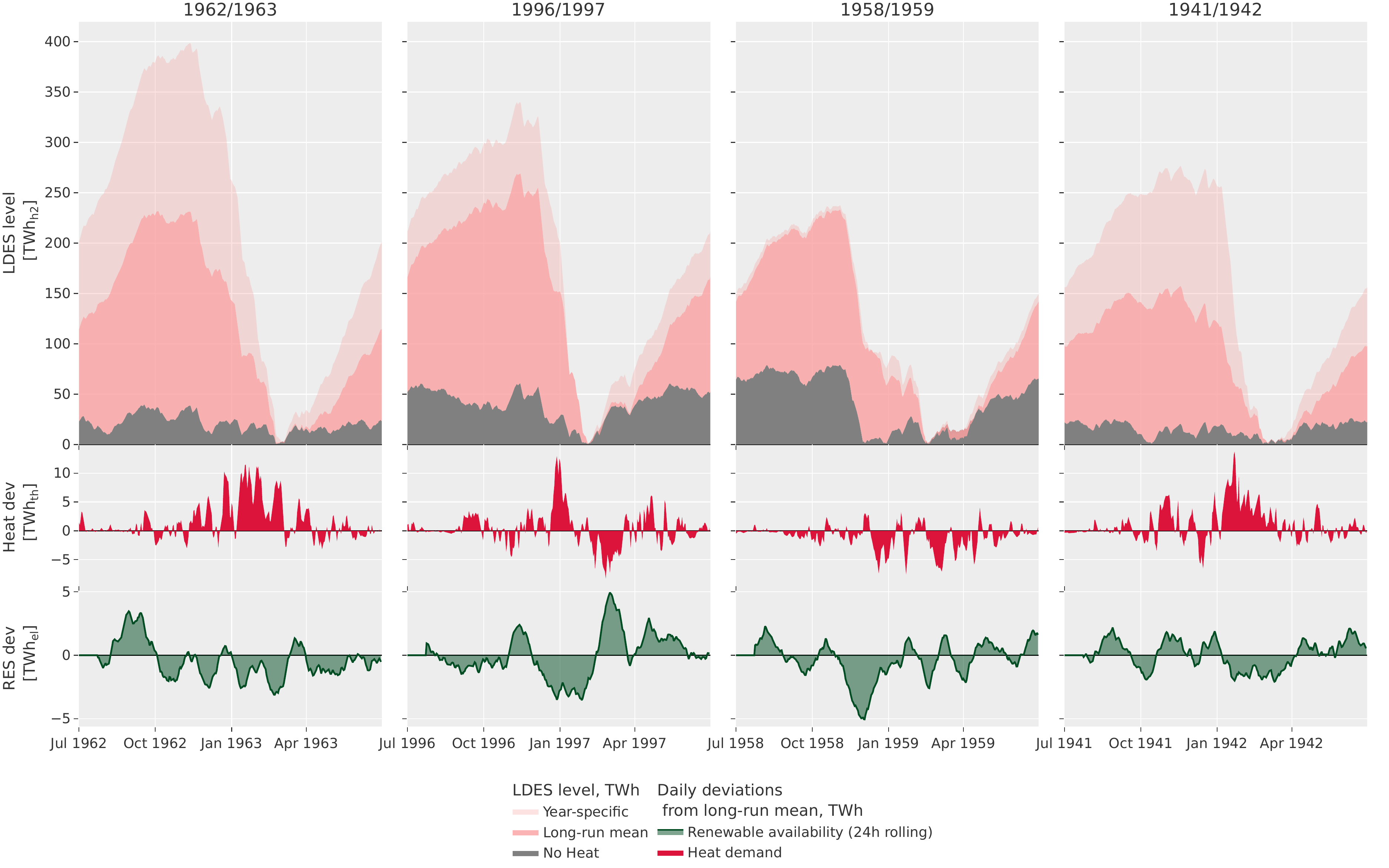}
    \label{fig:anecdotal}
\end{figure}

We illustrate both effects for four weather years with high additional \ac{LDES} requirements in Figure \ref{fig:anecdotal}. In the deterministic setting addressed here, the maximum required storage level in the course of a weather year determines the optimal \ac{LDES} energy capacity. The trajectories, therefore, provide an insight into the critical periods that determine optimal \ac{LDES} capacities. The gray areas correspond to the storage levels observed in the scenario without electrified heat (\textit{No Heat}). The darker red area reflects additional storage levels required to meet the mean heat demand profile (\textit{Decent - Mean}). The lighter red areas show the storage trajectories arising from the weather year's specific heat demand profile (\textit{Decent}). In the lower part of the figure, we plot the absolute daily deviations from long-run mean profiles of both heat demand (in dark red) and renewable generation. The heat deviations serve as a proxy for cold spells. For renewable generation, we take the mean generation portfolio of the \textit{Decent} scenario as a basis. The deviations are averaged in a 24-hour rolling window and are indicative of renewable energy scarcity, where the area under the zero line can be interpreted as an energy deficit.



We observe the most extreme \ac{LDES} capacity increase in 1962/63 (left). In the \textit{No Heat} scenario, the storage level traces four episodes of relative \ac{VRE} scarcity beginning in October 1962. The leverage effect amplifies all four episodes, particularly so in January when seasonal heat demand reaches its peak.
By some definitions, the winter of 1962/63 was the coldest European winter on record since 1739/40 \cite{greatbatch_tropical_2015} with a long, incessant cold spell from December to March encapsulating all renewable scarcity episodes and \textit{compounding} to a continuous discharge event of close to 334~TWh over the same period. 

The winter of 1996/97 (center left), often identified as an extreme weather year for renewable energy systems due to an episode of extremely low wind capacity factors \cite{kittel_quantifying_2024,grochowicz_using_2024}, was relatively mild, but included a brief cold snap in January. Featuring the second-highest storage requirements in the \textit{No Heat} scenario, it is a dominating leverage effect that adds \ac{LDES} capacity. The cold spell \textit{only} adds about 70~TWh.

The weather year of 1958/59 (center right) results in the highest \ac{LDES} capacity in the \textit{No Heat} scenario due to a severe episode of energy scarcity between November and early January, leading to a discharge event of 79~TWh over the course of 28~days. Rather mild temperatures induce a \ac{LDES} increase almost entirely through the \textit{leverage} effect, where the discharge event increases in size to 223~TWh and extends to 106~days. 

Lastly, the season of 1941/42 (right) exhibits one of the lowest \ac{LDES} requirements across all years in the sample when we do not include electrified heat. However, a moderate but prolonged \ac{VRE} scarcity period combined with a heavy cold spell in the first quarter of 1942 results in the fifth-highest \ac{LDES} requirements.

We note from these illustrative instances that different combinations of renewable scarcity events and cold spells can lead to significant \ac{LDES} capacity requirements.

\subsection{Long-duration thermal storage in district heating networks can mitigate \ac{LDES} requirements} \label{sec:mitigation}

So far, we have omitted large-scale \acf{LDTS} in district heating networks as an alternative storage technology. In the scenario \textit{District \& Decent}, we allow \ac{DH} networks to cover a share of the electrified heat and introduce large-scale \ac{LDTS}, which appear plausible only in such networks. \Ac{DH} network potentials vary by country (compare Figure \ref{fig:potentials}, left panel), with an average of around 32\% of total heat demand. We model \ac{LDTS} as pit thermal storage systems, a large insulated underground storage of hot water, connected to a \ac{DH} network. The \ac{DH} potential hence constrains the potential of \ac{LDTS} to replace \ac{LDES} capacities. We conservatively assume that \ac{DH} networks are only supplied by large-scale heat pumps, ignoring additional heat sources such as waste heat from hydrogen gas turbines or solar thermal collectors.

In contrast to a hydrogen cavern storage, an \ac{LDTS} has relatively high standing losses from heat loss \cite{sifnaios_heat_2023}. In the base specification, we assume an efficiency equivalent to cumulative losses of 39\% within 90~days of storage in energetic terms \cite{zeyen_mitigating_2021}. Hydrogen-based electricity storage, which includes electrolysis, cavern storage, and reconversion to electricity with gas turbines, has a relatively low roundtrip efficiency of around 30\%, but is independent of the storage duration as standing losses are negligible. Even under these conservative estimates for \ac{LDTS} efficiency, a \ac{LDTS} system is more efficient in terms of electrical energy input per unit of thermal energy demand served for storage duration of up to ca. 180~days~-~longer than the seasonal heating cycle.

\begin{figure}[H]
    \centering
    \includegraphics[width=\linewidth]{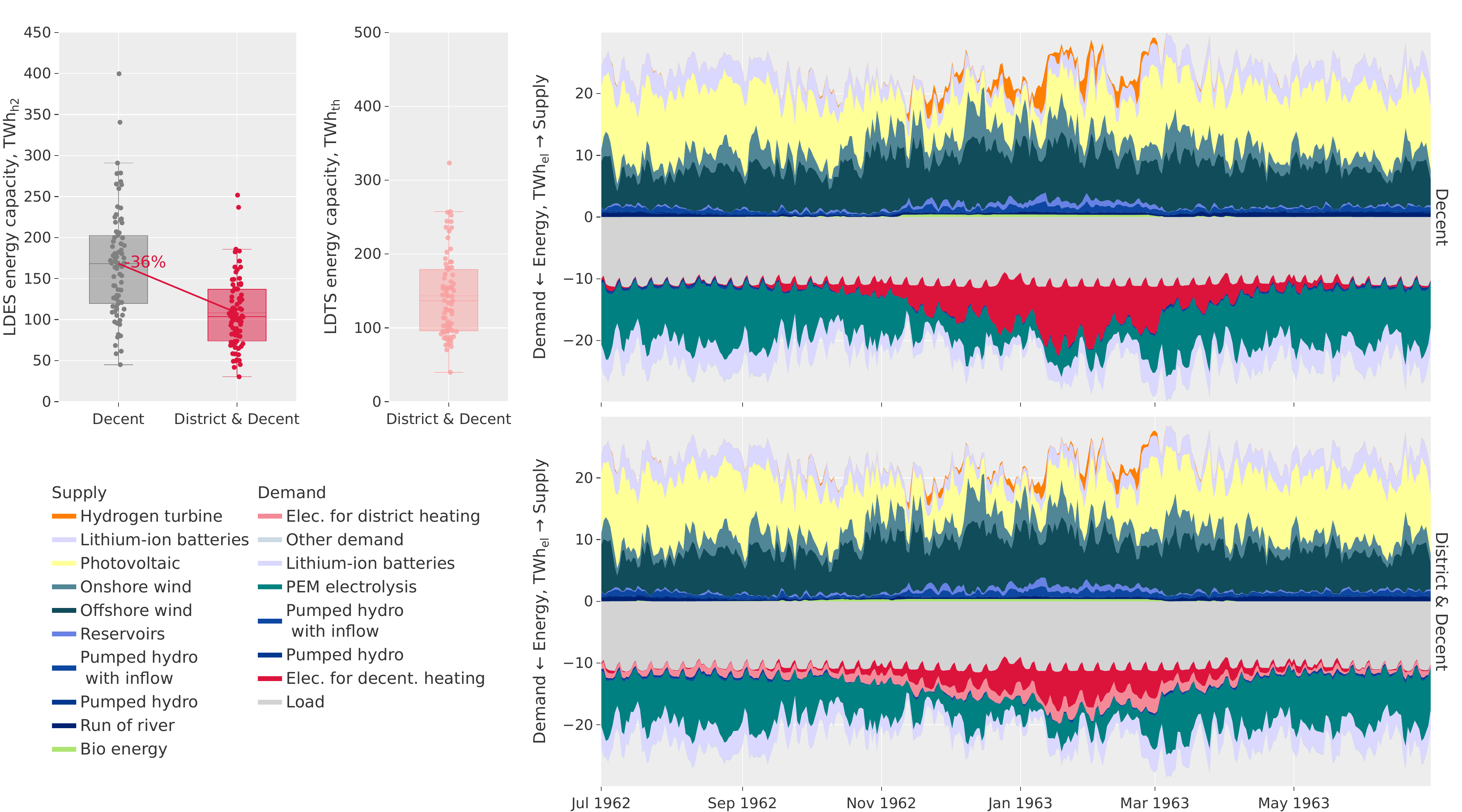}
    \caption{\textit{LDES requirements without and with district heating thermal storage} \\ \scriptsize{
        \textbf{left}: distribution of \ac{LDES} capacities across weather years under \textit{Decent} and \textit{District \& Decent} scenarios; \textbf{center}: distribution of \ac{LDTS} capacities in \textit{District \& Decent} scenario across weather years (center); \textbf{right}: 1962/63 daily electricity sector balance for \textit{Decent} scenario (upper) and \textit{District \& Decent} scenario (lower).
    }}
    \label{fig:storage_map}
\end{figure}

Our modeling results confirm such back-of-the-envelope calculations. The introduction of \ac{LDTS} reduces the additional \ac{LDES} requirements from electrified heat by 60~\ac{TWh} (36\%) on average (Figure \ref{fig:storage_map}, left panel). In the extreme weather year 1962/63, the introduction of co-optimized \ac{LDTS} capacities in \ac{DH} networks leads to a reduction of 155~\ac{TWh} (37\%). The Europe-wide average \ac{LDTS} capacity over all years is 143~TWh, reaching almost 323~TWh in the maximum (Figure \ref{fig:storage_map}, center panel). 

The right panel of Figure \ref{fig:storage_map} depicts the aggregate daily electricity sector balance for the year 1962/63 under \textit{Decent} (upper) and \textit{District \& Decent} (lower). Evidently, hydrogen turbine dispatch is significantly reduced in the \textit{District \& Decent} scenario as heating-network-connected \ac{LDTS} smooths and replaces electricity demand for heating. This effect is depicted by the reduced red area (electricity demand for heat pumps) in the figure.

Total electricity generation from hydrogen turbines decreases from 88~to~53~TWh on average (151~to~78~TWh for 1962/63). Peak generation falls from 65~to~34~GW (237~to~130~GW). Total electricity demand for heating increases slightly from 793 to 794~TWh on average (1038~to~1057~TWh) due to storage losses, but peak electricity demand from the heating sector falls from 351~to~280~GW (564~to~448~GW).  These findings hold for almost all weather years.

The introduction of \ac{LDTS} only slightly reduces generation capacity investments in solar PV (-5\% on average, -4.5\% for 1962/63) and wind (-1\% on average, -2\% for 1962/63), see Figure \ref{fig:impact_generation}.

\subsection{Central heat storage decentralizes overall long-duration storage infrastructure}

Potentials for cavern storage capacities are unevenly distributed across Europe (Figure \ref{fig:potentials}, left panel). Where there is potential in the form of rock salt formations in a country, the potential far exceeds any conceivably necessary storage requirement. Any of those countries could technically host all the cavern storage required for the entire continent. Yet, in the \textit{Decent} scenario, around 43\% (72 out of 168~TWh) of the total cavern storage capacity is located in Germany on average, compare Figure \ref{fig:geog_decentral}. Notably, cavern storage is much more evenly distributed between countries in the \textit{No Heat} scenario (Figure \ref{fig:geog_no_heat}, left panel). Hydrogen gas turbines for reconversion are also concentrated in Germany in \textit{Decent}, while electrolysis capacity is distributed more evenly, across countries on average.

\begin{figure}[H]
    \centering
    \includegraphics[width=\linewidth]{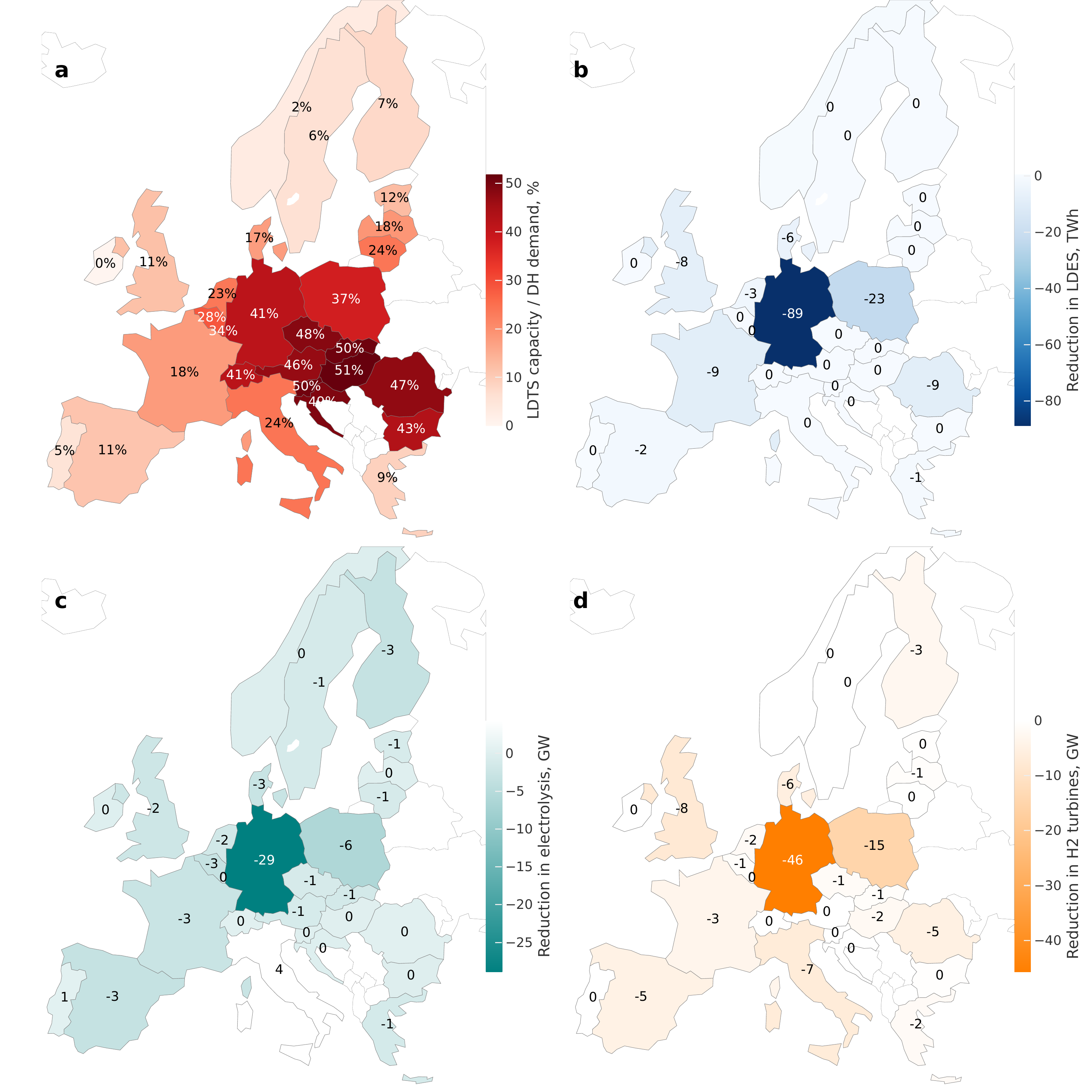}
    \caption{\textit{Geographical impact of adding long-duration thermal storage in 1962/63} \\ \scriptsize{
    (a) Share of district heating heat demand covered by LDTS; (b) Change in LDES capacity when introducing LDTS; (c) Change in PEM electrolysis when introducing LDTS; (d) Change in H2 turbine capacity when introducing LDTS.}}
    \label{fig:geogr}
\end{figure}

Figure \ref{fig:geogr} shows the change in the geographical long-duration storage distribution across Europe when district heating networks with \ac{LDTS} are introduced for the example of 1962/63. Heating demands and district heating potentials vary between countries, such that a meaningful comparison of \ac{LDTS} deployment requires a relative basis. We express \ac{LDTS} as a share of total demand in district heating networks. \ac{LDTS} uptake is particularly high in Central Eastern Europe where there is very little \ac{LDES} capacity. The reduction in \ac{LDES} capacity is disproportionately high in Germany. Consequently, the share of European \ac{LDES} capacity in Germany falls to  35\% (38 out of 108~TWh) in \textit{District \& Decent}. This masks two effects. For one, Germany has a relatively high district heating potential compared to other cavern storage locations, which means that it can substitute some of its on flexibility requirements locally. For another, German \ac{LDES} provides flexibility in other countries, particularly in Central Eastern Europe and high \ac{LDTS} adoption in these countries also reduces German \ac{LDES} requirements.

\subsection{Sensitivities} \label{sec:sens}

We investigate a range of sensitivities for the results. For computational tractability, we limit these runs to the four years discussed above. 

\begin{figure}[H]
    \centering
    \includegraphics[width=\linewidth]{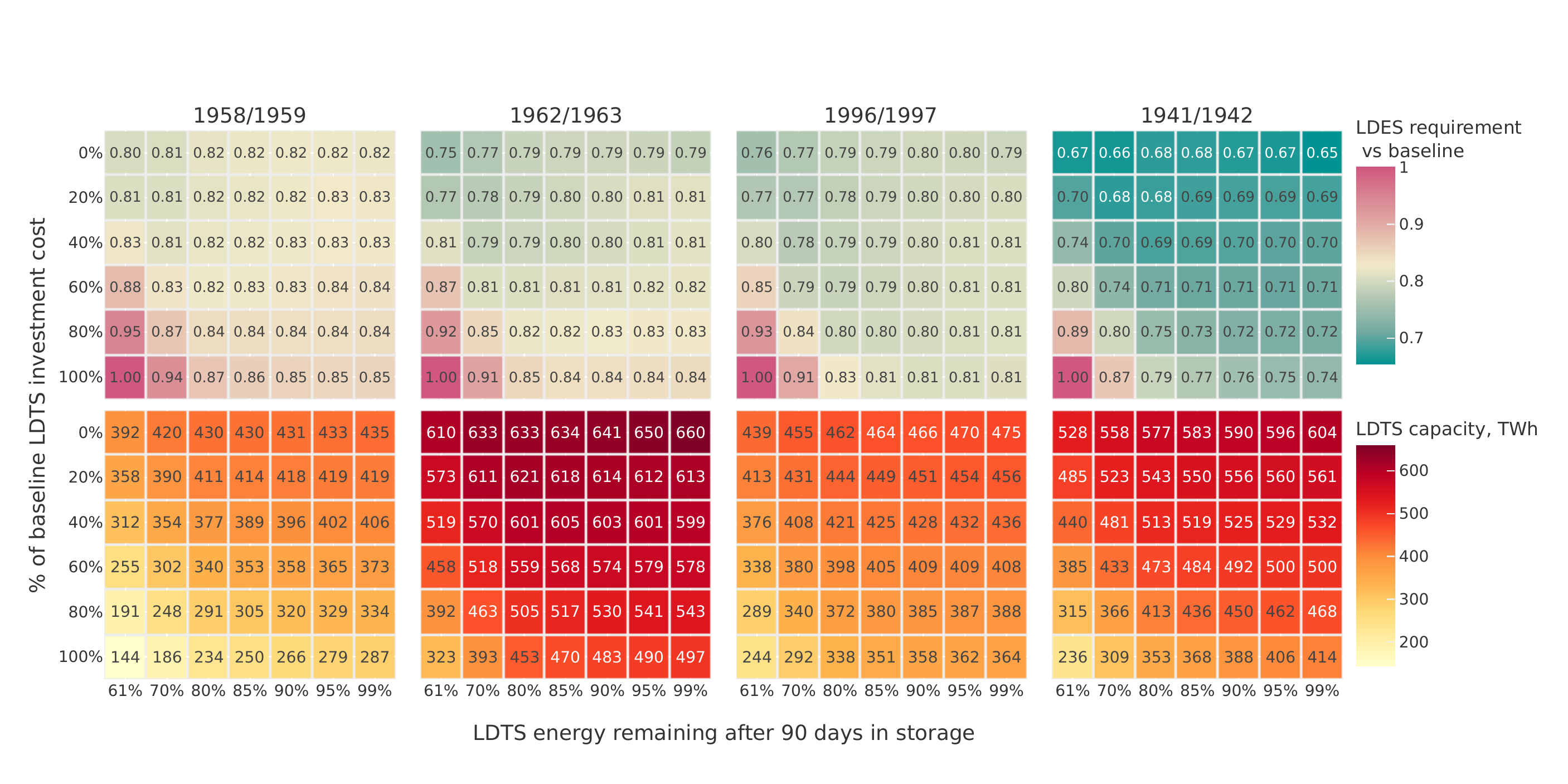}
    \caption{\textit{LDTS sensitivity results} \\ \scriptsize{
    Sensitivity of \ac{LDTS} parameters. The x-axis varies heat losses from the base case to almost no losses in energetic terms, expressed as energy remaining after 90 days in storage. The y-axis varies the assumed investment costs of LDTS by multiplying them by the labeled scalers.}}
    \label{fig:heatmap}
\end{figure}

We test alternative specifications of \ac{LDTS} efficiencies and investment costs. For each of the four weather years, we run 42 alternative settings, varying the \ac{LDTS} standing losses on the x-axis and the investment costs on the y-axis (Figure \ref{fig:heatmap}). The base configuration used in all prior scenarios above is in the bottom-left corner. In the most favorable parameter configurations \ac{LDTS} reduces \ac{LDES} needs by another 19-35\% (upper panel). In the case of 1941/42, this would correspond to an \ac{LDES} requirement of 111~TWh when \ac{LDTS} is free and efficiency is at 99\%. As per the lower panel of Figure \ref{fig:heatmap}, such additional reductions are accompanied by large increases in optimal \ac{LDTS} capacities. For 1941/42, the \ac{LDTS} capacity almost triples between the base case and the best parameter combination. Based on the conclusion above that even in the base specification, heat losses do not match the round-trip losses of \ac{LDES} for relevant storage durations, it is not surprising that additional substitution from efficiency gains beyond a threshold is limited. In the top line of each tile, \ac{LDTS} can be built for free. In these cases, the substitution potential with respect to \ac{LDES} is limited by the country-level \ac{DH} potentials.

Next, we explore an extreme counterfactual case in which all electrified heating would be supplied through heating networks (Table \ref{tab:sens}, last row), which makes \ac{LDTS} available for all electrified heating demand. In such a setting, \ac{LDES} requirements would be significantly reduced, but still remain above the \textit{No Heat} scenario. This reflects the fact that \ac{LDTS} smooths heating electricity demand peaks by shifting part of the load to months outside the heating season, but does not eliminate all additional electricity demand in winter. This still leaves a \textit{leverage} effect (compare right panel of Figure \ref{fig:storage_map}), as described in Section \ref{sec:lev_comp}.

\begin{table}[H]


\caption{\textit{Sensitivities} \\ \scriptsize{
Aggregate European \ac{LDES} requirements in TWh for selected sensitivities. 'Base case' refers to the scenarios used in the remaining analysis. 'No H$_2$ network' assumes that there is no hydrogen network. 'Islands' assumes that there is no interconnection between countries at all, 'Nuclear' assumes fixed nuclear generation capacities of 96.9~GW \cite{entso-e_tyndp_2025}. 'All district heating' assumes that all electrified heating can be supplied in heating networks with \ac{LDTS}.}}

\input{tables/sensitivities}

\label{tab:sens}
\end{table}

We also test the impact of different interconnection settings. First, we assume that there is no hydrogen network, see rows 3 and 7 of Table \ref{tab:sens}. Here, the increase in \ac{LDES} energy capacities is more pronounced in the \textit{District \& Decent} scenario (20\% - 44\%) than in the \textit{Decent} scenario (8\% - 32\%). In the \textit{District \& Decent} base case, hydrogen imports with long-term contracts are actually slightly higher than in the \textit{Decent} base case, which is likely to replace some \ac{LDES} capacity otherwise used for serving flat industrial hydrogen demand. In the absence of hydrogen import capacities, this option no longer exists.

In the scenario with neither hydrogen nor electricity interconnection capacity ('Islands', rows 4 and 8), \ac{LDES} capacities increase only marginally compared to the 'No H$_2$ network' case and even decrease in the case of 1996/97 in the \textit{Decent} scenario, indicating that the system value of \ac{LDES} depends on interconnection. Without interconnection, building larger generation portfolios is cheaper than building expensive hydrogen tank storage. Only about 10~TWh of tank storage is being built across Europe in the 'Islands' case of the  \textit{Decent} scenario.

Lastly, we test the effect of introducing around 97~GW of nuclear capacity, of which the majority is located in France (58~GW). In either scenario, it has a limited effect on \ac{LDES} capacities (-14\% and -18\% on average). The reason is likely to be twofold. For one, nearly two-thirds of the capacity is in France, and limited interconnection capacities may prevent spatial balancing during extreme system episodes. By contrast, \ac{LDES} leaves the option to transport both electricity and hydrogen. For two, as Section~\ref{sec:mitigation} shows, peak supply by hydrogen turbines in winter periods clearly exceeds the aggregate nuclear capacity in extreme years, and hence the introduced nuclear capacities can only supply a limited part of the positive residual load.

\section{Discussion and conclusion} 

\subsection{Summary}

In this analysis, we isolate the effect of the electrification of heating with heat pumps on optimal long-duration storage needs in a fully renewable European energy system using a large sample of 78 weather years. We find that heat pumps significantly increase \acf{LDES} requirements, mainly because additional electricity demand has to be served in periods of renewable energy scarcity, which typically occur in the heating season during winter \cite{mockert_meteorological_2023,kittel_quantifying_2024}. This \textit{leverage} effect accounts for 75\% of the average additional \acf{LDES} requirements. The remaining 25\% can be explained by a \textit{compound} effect, describing the coincidence of the aforementioned periods of renewable energy scarcity and cold spells, which are periods of exceptionally high heating demand. This leads to considerable positive residual load events that have to be covered by \ac{LDES}. Similar compound effects are found in previous analyses of extreme system states \cite{gotske_designing_2024,mockert_meteorological_2023}.


Previous literature on \ac{LDES} has noted the sensitivity of optimal \ac{LDES} capacities to interannual weather variability \cite{dowling_role_2020,ruhnau_storage_2022,kittel_coping_2025}. Yet, these studies have only considered supply-side variability across weather years. Our results suggest varying load, driven by temperature-dependent heating demand, has a much greater impact on optimal \ac{LDES} deployment. The impact of demand-side weather variability further highlights the necessity to include large datasets in energy system models when planning weather-resilient renewable energy systems.

The use of \acf{LDTS} can effectively reduce the need for \ac{LDES} but is likely confined to district heating networks. Using an optimistic estimate of the potential for future district heating networks, we find that even a conservatively parameterized \ac{LDTS} can substitute up to 36\% of the \ac{LDES} requirement in a least-cost system.

While the long-duration storage infrastructure is concentrated in Germany in a scenario with only decentralized heat pumps, the introduction of \ac{LDTS} 
leads to a more even geographic distribution of storage capacities. With its limited hydrogen cavern potential, \ac{LDTS} in Central Eastern Europe can relieve some cavern storage needs in Germany. 

Our findings are fairly robust to a range of sensitivities. Cost reductions and efficiency gains for \ac{LDTS} would further mitigate \ac{LDES} requirements. Notably, policy-relevant capacities of nuclear power in Europe mitigate \ac{LDES} and \ac{LDTS} needs only to a limited extent. Likewise, alternative assumptions on interconnection capacities have only minor impacts on results. For very optimistic alternative assumptions on the costs, efficiency and potentials of \ac{LDTS}, we find that thermal storage mitigates hydrogen storage needs to a much larger extent than under baseline assumptions.

\subsection{Limitations}

The large geographic and temporal scope of our analysis comes at the expense of simplifications in the representation of several model components. Specifically, we only use one type of air-sourced heat pump for decentralized heating and do not include any efficiency improvements in the building stock. Similarly, we aggregate all district heating networks within a country and only consider one large-scale heat pump to supply these, abstracting from other potential heat sources. We further use a simplified representation of the thermodynamics of large-scale pit thermal storage, which neglects that some electricity might be required at the time of \ac{LDTS} discharging to elevate the temperature level again, which could reduce the cost advantage of thermal storage against hydrogen storage. While some of these assumptions (e.g., only air-sourced heat pumps, no other heat sources) lead to a higher electricity demand and higher \ac{LDES} needs than expected in reality, others might lead to an underestimation (simplified representation of the thermodynamics). Finally, the perfect foresight assumption for a given weather year leads to ideal but unrealistic storage operations and is likely to underestimate the relative value of \ac{LDES}.

\subsection{Conclusions \& outlook}

In our most challenging setting with 80\% electrified heat demand, in the absence of long-duration thermal storage, and considering an extremely cold and long winter such as 1962/63, we find a European \acf{LDES} capacity requirement of 400~TWh. While this capacity is very large compared to currently installed electricity storage, it appears technically feasible considering geological potentials for caverns in Europe \cite{caglayan_technical_2020}. For context, the currently existing natural gas storage capacity in the EU is around 1,100~TWh \cite{agency_for_the_cooperation_of_energy_regulators_report_2022}. However, only a part of the existing gas caverns are suitable for hydrogen storage due to the more demanding geological requirements \cite{oni_underground_2025} and the lower volumetric density of hydrogen compared to natural gas reduces the energy storage capacity. Presently, bar a few test sites, hydrogen cavern storage is a nascent technology.

Long permitting processes and underdeveloped supply chains, as well as long-term uncertainty about hydrogen prices and quantities, are likely to increase project risks \cite{steinbach_future_2024}. Similarly, neither electrolysis nor hydrogen gas turbines have been deployed at scale so far \cite{odenweller_green_2025}. Moreover, uncertainties around the ramp-up of heat electrification and hydrogen transmission infrastructure complicate the economics of highly centralized \acf{LDES}. In this context, \ac{LDTS} could provide a promising complementary technology, as it does not rely on particular geological conditions and can effectively deal with long-duration flexibility requirements in the heating sector. Yet, it is likely to be confined to heating networks, and our results suggest that even if they were substantially expanded, a sizable residual long-duration flexibility need from the heating sector would remain. 

Our results indicate that \ac{LDES} in the form of cavern storage would be concentrated in a few countries, particularly in Germany, and thus other countries without salt cavern potentials to store hydrogen would be dependent on imports and transmission infrastructure to cover their load and heat demand. As this could raise regulatory and geopolitical questions, \ac{LDTS} could, for some countries, increase their energy autonomy and decrease dependence on energy transport infrastructure.

Accordingly, \ac{LDES} and \ac{LDTS} should not be considered as substitutes, but rather complement each other, and policymakers should work to enable the mass deployment of both storage types. In particular, policymakers and regulators should expedite the process of designing adequate regulatory frameworks that enable a timely capacity build-out, as a lack of long-duration flexibility would jeopardize the electrification of the heating sector. 

From an investor's perspective, the increased variance of optimal \ac{LDES} capacities between weather years, when considering electrified heating, raises additional questions around predictable cost recovery without other economic incentives. This clearly calls for de-risking such investments. Compared to hydrogen storage, it might be easier to manage such risks for thermal storage, as the latter is likely to be developed and operated by vertically integrated, cost-regulated regional monopolists who should be better able to manage quantity and price risks.

Our results show that not only supply-side, but also demand-side weather dependency has a sizable effect on renewable energy systems. Energy modelers and system planners should take this into account when investigating weather-resilient renewable energy systems. To fully capture not only the leverage but also the compound effect, as many weather years as possible should be considered in respective analyses. 

As for future research, we consider a more granular representation of the heating sector in combination with a finer geographical resolution a promising avenue. In particular, including alternative heat supply sources such as solar thermal or waste heat from hydrogen turbines or other sources could impact the optimal shares of centralized \ac{LDES} and local \ac{LDTS}. Building renovation and the trade-off with storage \cite{zeyen_mitigating_2021} should be appraised in light of interannual weather variability. 

Finally, the use of deterministic models omits a potentially important advantage of \ac{LDES} over \ac{LDTS} in settings without perfect foresight. \ac{LDES} can serve final energy demand in many different sectors, depending on the realized system state, thanks to the usefulness and convertibility of hydrogen. In contrast, \ac{LDTS} is less useful to the energy system as it can only serve heat demand. Future research, using stochastic modeling (such as in Schmidt \cite{schmidt_long-duration_2025}) could be a starting point to better understand the value of different types of storage, the interaction between them, and their operation in a non-deterministic setting. However, such approaches are likely to be feasible only for less extensive geographical and temporal scopes than modeled here.

\section{Methods} \label{sec:methods}

This paper uses a sector-coupled capacity expansion model of Europe across a sample of 78 weather years (1940/41 - 2017/18) at hourly resolution in a scenario analysis to isolate the effect of electrified heating on long-duration storage requirements.

\subsection{Sector-coupled energy system model}

The energy system model used is an extended version of the open-source power system model DIETER \cite{zerrahn_long-run_2017}, written in Julia. The model has been used in various studies to investigate flexibility in sector-coupled \ac{VRE}-based energy systems, e.g. \cite{zerrahn_economics_2018,schill_electricity_2020,roth_geographical_2023,roth_power_2024,gueret_impacts_2024}. DIETER is a linear cost-minimizing model that abstracts from discrete capacity decisions and economies of scale as well as non-convex operational constraints. The model is centered around the power sector, modeling the target year 2050.

\subsubsection{Geographical scope and grid assumptions}

The model includes a network of 28 countries (EU27 excluding Malta and Cyprus plus Norway, the United Kingdom and Switzerland). Transmission network capacities are on a net transfer (NTC) basis and fixed to the TYNDP 2022 Reference Grid \cite{entso-e_tyndp_2022} (Figure \ref{fig:networks}, left panel).

\subsubsection{Generation technologies}

The model only permits renewable generation technologies including solar photovoltaics, onshore and offshore wind, hydro generation and some biomass combustion as well as gas turbines running on hydrogen. We only admit nuclear generation capacity in a sensitivity (see Section \ref{sec:sens}). We assume high technical capacity potentials for solar and onshore wind \cite{trondle_home-made_2019}, see Table \ref{tab:capa_bounds}. Offshore wind capacities are additionally constrained to remain in the range of the Ten-Year Network Development Plan (TYNDP 2024) of the European Network of Transmission System Operators for Electricity (ENTSO-E) \cite{entso-e_tyndp_2025}. Biomass capacities are to remain at most at today's levels and total annual generation is constrained to that of 2022 \cite{ember_yearly_2024}. Hydro capacities, including reservoirs and run-of-river, are fixed at today's levels. Today's capacities are based on the European Resource Adequacy Assessment (ERAA) \cite{de_felice_entso-e_2022}. 

\subsubsection{Storage technologies}

Storage technology candidates are limited to lithium-ion batteries and pumped-hydro storage (PHS) for shorter periods. Long-duration storage options are hydrogen tanks, hydrogen caverns, and pit thermal energy storage, described in more detail below. The model chooses battery capacities endogenously, while PHS is fixed to today's levels \cite{de_felice_entso-e_2022}, see  Table \ref{tab:hydro_capas}.

\subsubsection{Demand}

The demand side of the power system balance consists of an exogenous and an endogenous part. The exogenous part features a historical base electricity demand time series from Open Power System Data \cite{muehlenpfordt_open_2020} that has been corrected for existing electrified heat demand using data from When2Heat \cite{ruhnau_update_2022}. We add incremental industrial electricity demand resulting from projected process electrification until 2050 \cite{neumann_potential_2023}, assuming a flat profile and, for computational reasons, we also assume a flat demand profile to represent electric vehicle demand in 2050 based on the size of today's car fleet. The implicit assumption is that the short-term variations in mobility demand and the flexibility potential of electric vehicles do not interact substantially with long-duration storage requirements beyond the load they add to the system. The left panel of Figure \ref{fig:energy_demand} gives the resulting (exogenous) annual electricity demand by country.

The endogenous electricity demand comes from storage technologies, hydrogen electrolysis and heat pumps, either decentrally or from large-scale heat pumps connected to district heating networks. A minor share of electricity demand comes from compressors in the hydrogen network, as explained below. 

Apart from an electricity balance, the model also imposes balances for hydrogen and decentral heat as well as district heating networks. 

\subsubsection{Hydrogen}

A country's hydrogen supply can either be met by domestic production using proton exchange membrane (PEM) electrolysis, imports from other countries within, or imports from countries outside the scope of the model. In the case of the latter, imports can arrive by ship or by pipeline. Such external imports are assumed to be possible only under long-term contracts (LTC) that provide a baseload supply with some hourly flexibility of +/- 10\%. We assume that shipped hydrogen comes in as cracked ammonia \cite{spatolisano_ammonia_2023}. Import and interconnection capacities are taken from TYNDP 2024, see Figure \ref{fig:networks}, right panel. We assume LTC import prices of 86.8~€/MWh for shipped hydrogen (ammonia + cracking) and 42-51~€/MWh for pipeline imports \cite{entso-e_tyndp_2025}. To use the network, hydrogen generated by electrolysis needs to be compressed, which creates electricity demand. Longer transport distances require booster compression, resulting in additional electricity demand. 

Hydrogen is stored in either salt caverns underground or more expensive pressurized tanks. For both storage options, additional compression is necessary. Cavern storage potentials by country are taken from \cite{caglayan_technical_2020} (see Figure \ref{fig:potentials}, right panel). Industrial demand for hydrogen is assumed to have a baseload profile, and annual quantities are taken from TYNDP~2024 (see Figure \ref{fig:energy_demand}, middle panel). We assume that all other uses of hydrogen, and especially its derivatives, will be covered by additional imports and do not enter the hydrogen balance. Another source of hydrogen demand stems from hydrogen gas turbines that combust hydrogen to produce electricity. The combination of electrolysis, cavern storage and hydrogen turbines forms the \acf{LDES} discussed in this paper.

\subsubsection{Heating}

The model includes the share of the heating sector that is assumed to be electrified with heat pumps. We assume the use of air-sourced heat pumps. The efficiency of heat pumps varies with the outside temperature, see Figure \ref{fig:input_heat}, lower panel. For decentralized heat pumps, we set the heat pump capacity exogenously to meet the peak heat demand. We assume that decentralized heat pumps are equipped with a small buffer storage with a duration of 1.5~hours \cite{roth_power_2024}.

District heating networks are modeled as one large energy balance by country. The energy supply is restricted to large-scale air-sourced heat pumps, for which the model chooses optimal capacities. \Acf{LDTS} is modeled as pit thermal energy storage, large, water-filled excavated ground sites insulated with waterproof liners. Several dozen sites are already in operation across Europe \cite{xiang_comprehensive_2022}. Pit thermal storage is subject to heat loss. We assume a conservatively high loss, leaving 61\% of the energy after 90~days in storage \cite{zeyen_mitigating_2021}. Large-scale pit storages like Dronninglund in Denmark have shown higher efficiencies in long-run investigations \cite{pan_long-term_2022}. We further assume a \ac{LDTS} (dis-)charge efficiency of 90~\% to account for heat loss in district heating network pipes. 

\subsection{Data sources}

\subsubsection{Cost parameters}

Most cost parameters are taken from the Danish Energy Agency's (DEA) forecasts for 2050 \cite{danish_energy_agency_technology_2024}. We provide a detailed list in Table \ref{tab:cost_assumptions}. All costs and prices are in 2022€. The value of lost load is set at 100,000~€/MWh.

\subsubsection{Weather data} \label{sec:weather_data}

We use the largest complete dataset of weather inputs for energy system models that is openly available at the time of writing \cite{antonini_weather-_2024}. We use data from 1940 to 2018. The dataset includes country-level capacity factors for solar photovoltaics, wind onshore and wind offshore as well as inflow data for run-of-river, reservoirs and open pumped-hydro storage. Importantly, it also includes residential and commercial heat demand data for all European countries. The data is based on the ERA-5 reanalysis data of the Copernicus project \cite{hersback_era5_2023} and we use temperature data from the same dataset to generate the coefficient of performance time series using the When2Heat tool \cite{ruhnau_update_2022}. We have corrected heat demand and hydro time series for long-run trends \cite{gotske_designing_2024}. We provide details on our approach in SI.XX. Finally, we scale heat demand by annual data from \cite{rozsai_m_jrc-idees-2021_2024}. The resulting annual demands are shown in Figure \ref{fig:energy_demand}, right panel. We note that total demands and implicit heating thresholds reflect today's building stock. Large-scale insulation and retrofitting could significantly reduce effective useful heating demand in 2050 \cite{zeyen_mitigating_2021}. 

For the scenarios with central heating networks, we apply the country-wise shares of district heating of total heat demand projected in an optimistic scenario for 2050 by \cite{fallahnejad_district_2024} (Figure \ref{fig:potentials}, left panel). For each hour, we apply these shares to the hourly heat demand and assume that the remainder up to 80\% is to be supplied by decentralized heat pumps, effectively creating two heat demand time series per country. 

\subsection{Scenario design}\label{sec:scenario_design}

We investigate the role of electrified heating for long-duration storage requirements using four main scenarios. Table \ref{tab:scenarios} provides descriptions for each of them.


We assume that 80\% of space and water heat demand in residential and commercial buildings (not industry) will be covered with electrified heating options, namely decentralized small-scale heat pumps and large-scale heat pumps within district heating networks. These assumptions are conservative in some respects and progressive in others. 

The assumed share of 80\% is surely at the upper end of the range of projected electrification shares. A study by the Joint Research Center (JRC) of the European Commission \cite{tsiropoulos_towards_2020} gives an overview of the final energy consumption in the EU28 buildings sector of different scenarios that reach at least 90\% emission reduction by 2050. The share of electricity and ambient ranges around 60\% for most scenarios, yet reaches up to 80\% for others. The leading German energy transition policy research consortium considers various scenarios for the building sector in Germany, in the majority of which decentralized heat pumps and district heating cover more than 80\% of the total heat demand by 2045 \cite{luderer_energiewende_2025}. 

At the same time, heat electrification only based on heat pumps is likely to be conservative in terms of the demands on the power sector, as it is more efficient than alternatives such as resistive heaters or even hydrogen boilers \cite{rosenow_meta-review_2024}.

\section*{Data availability}

\noindent Results data are curated in a Zenodo repository (\href{https://zenodo.org/records/15482638}{https://zenodo.org/records/15482638}). Technology input data can be reproduced by an auxiliary repository (\href{https://gitlab.com/diw-evu/dieter_public/dieterdata}{https://gitlab.com/diw-evu/dieter\_public/dieterdata}). The weather input data is compiled in the Gitlab repository linked below.


\section*{Code availability}
\noindent The project code is openly available in a Gitlab repository (\href{https://gitlab.com/diw-evu/projects/long-thermal}{https://gitlab.com/diw-evu/projects/long-thermal}). The analysis is written in Julia and Python and the README file explains the intended order of execution.


\section*{Acknowledgments}
\noindent We acknowledge a research grant from the German Federal Ministry of Education and Research via the
Ariadne projects (FKZ 03SFK5NO-2).


\section*{CRediT author contributions}

\noindent\textbf{Felix Schmidt}: Conceptualization, methodology, model development, software, analysis, data curation, visualization, writing — original draft, review and editing 

\noindent\textbf{Alexander Roth}: Conceptualization, methodology, analysis, writing — original draft, review and editing 

\noindent\textbf{Wolf-Peter Schill}: Conceptualization, methodology, analysis, writing — review and editing 



\newpage
\printbibliography



\newpage
\appendix
\setcounter{figure}{0}
\renewcommand{\thefigure}{SI.\arabic{figure}}
\setcounter{table}{0}
\renewcommand{\thetable}{SI.\arabic{table}}
\renewcommand{\thesubsection}{SI.\arabic{subsection}}

\section*{Supplemental Information}

\subsection{Additional model output} \label{sec:additional_output}
\subsubsection{Generation portfolio}

\begin{figure}[H]
    \centering
    \includegraphics[width=\linewidth]{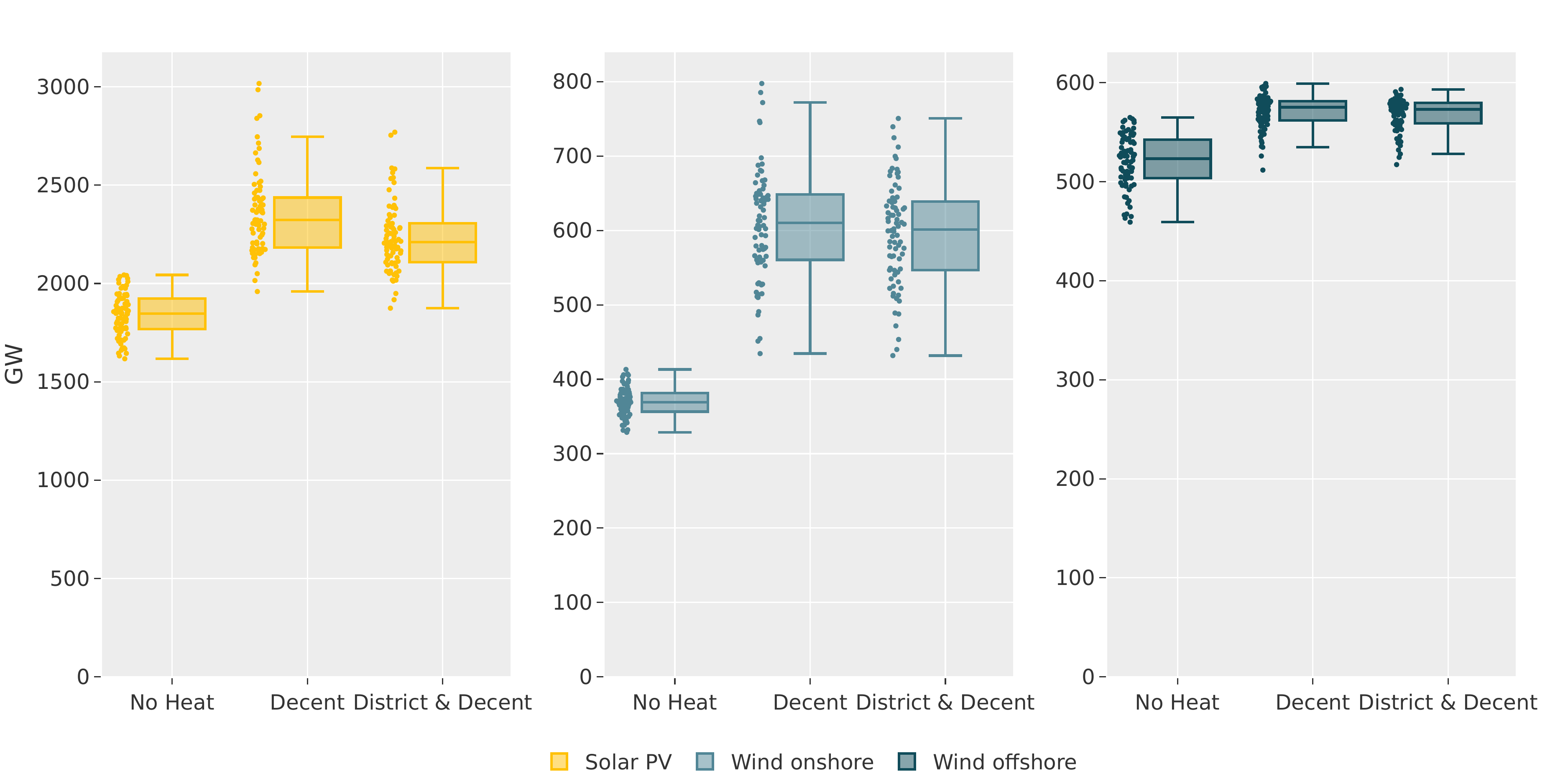}
    \caption{\textit{Distribution of aggregate renewable generation capacities under different scenarios.} }
    \label{fig:impact_generation}
\end{figure}

\subsubsection{Spatial dimension}
\begin{figure}[H]
    \centering
    \includegraphics[width=\linewidth]{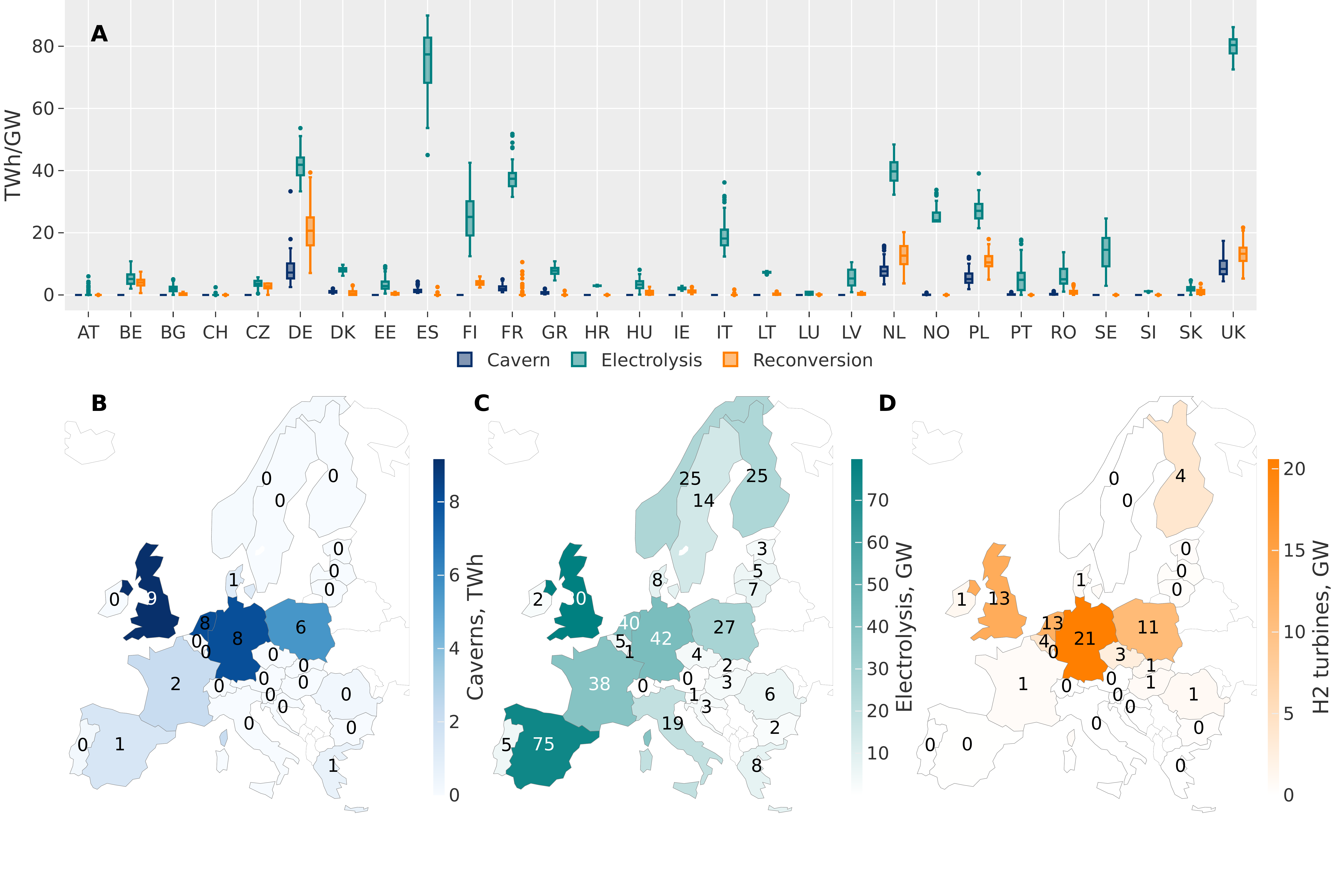}
    \caption{\textit{Geographical distribution of LDES capacities - No Heat}: (A) Distribution of cavern, electrolysis and h2 turbine capacities by country across weather years (1940/41 - 2017/18), (B) - (D) Mean capacities }
    \label{fig:geog_no_heat}
\end{figure}
\begin{figure}[H]
    \centering
    \includegraphics[width=\linewidth]{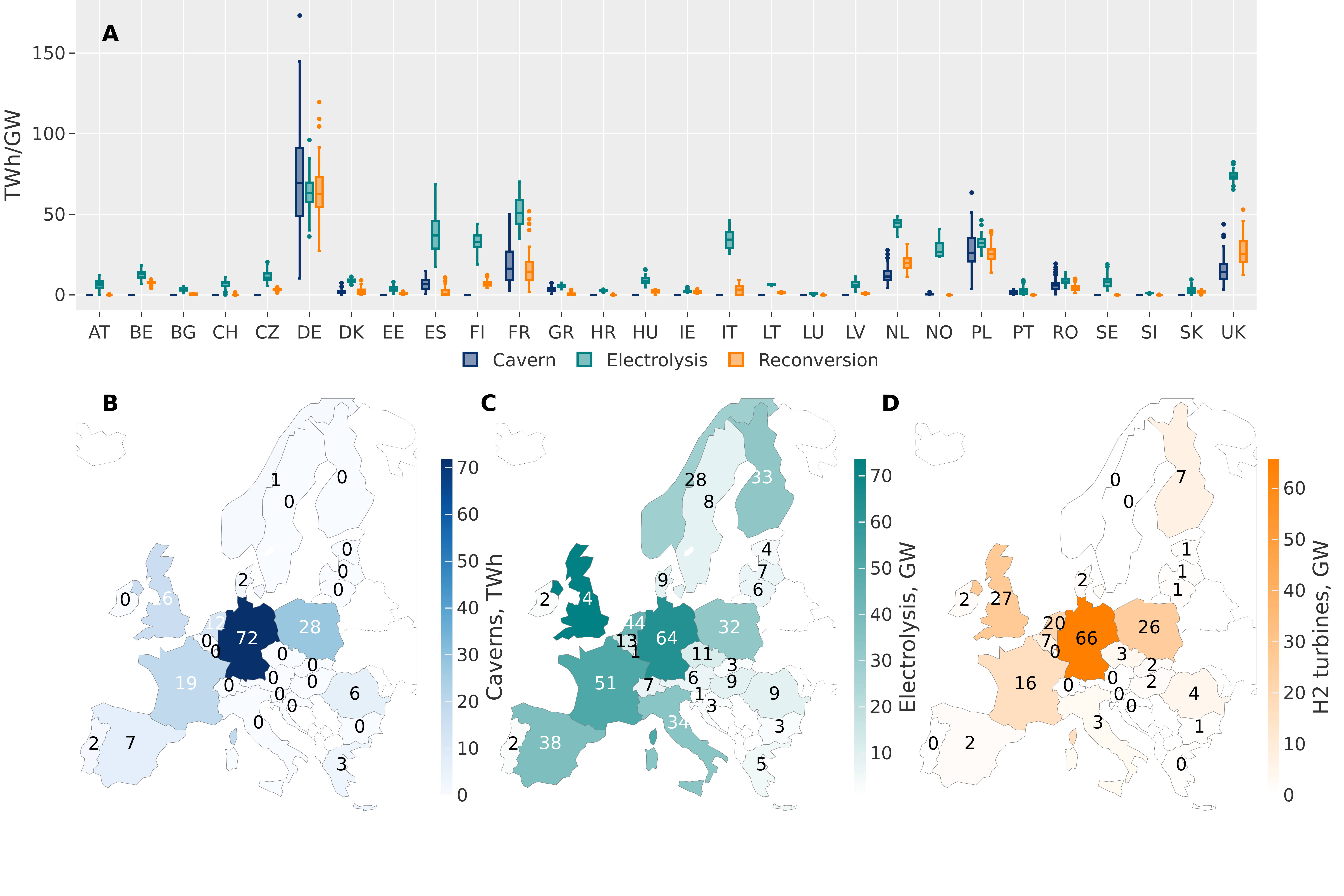}
    \caption{\textit{Geographical distribution of LDES capacities - Decent}: (A) Distribution of cavern, electrolysis and h2 turbine capacities by country across weather years (1940/41 - 2017/18), (B) - (D) Mean capacities }
    \label{fig:geog_decentral}
\end{figure}
\begin{figure}[H]
    \centering
    \includegraphics[width=\linewidth]{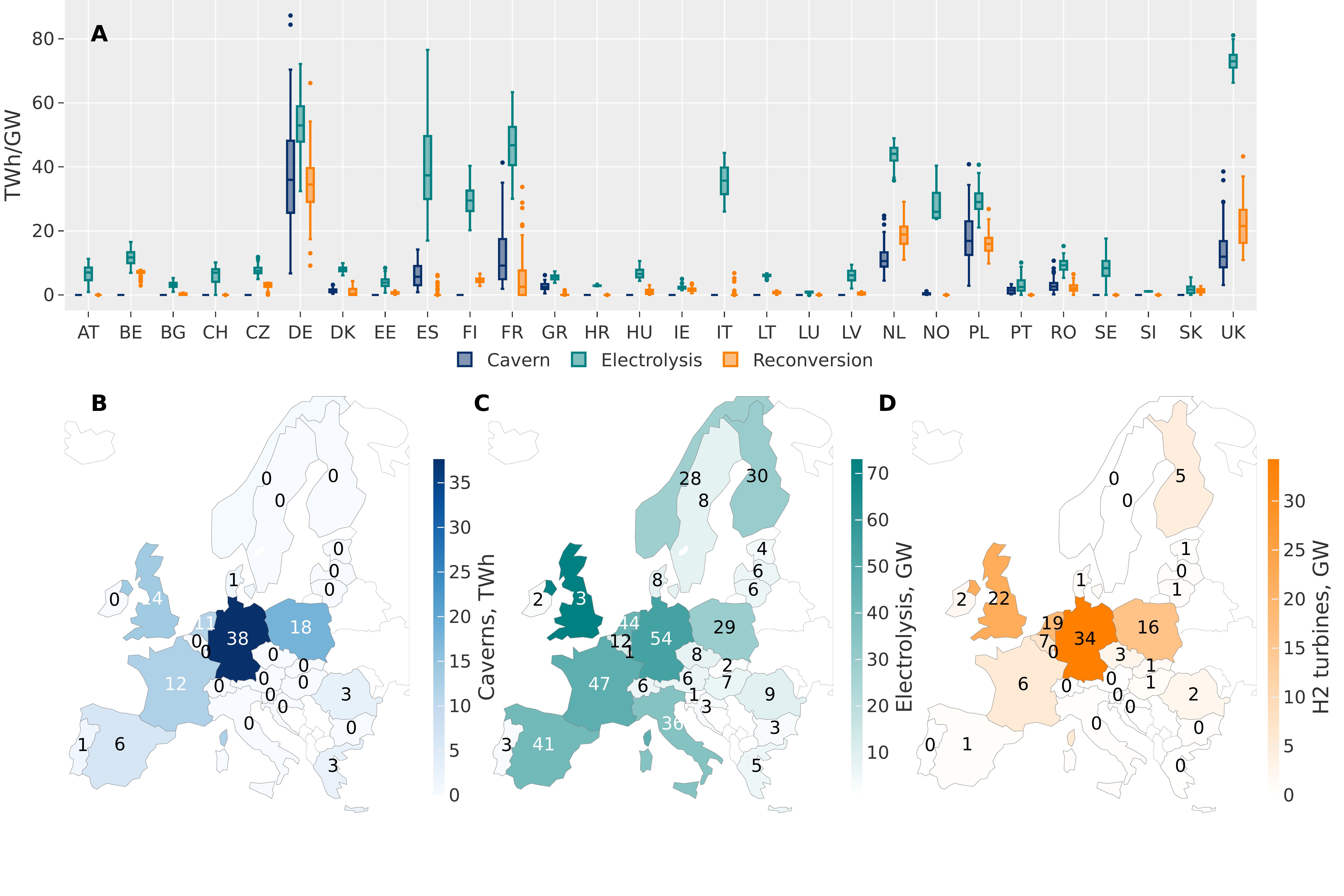}
    \caption{\textit{Geographical distribution of LDES capacities - District \& Decent}: (A) Distribution of cavern, electrolysis and h2 turbine capacities by country across weather years (1940/41 - 2017/18), (B) - (D) Mean capacities }
    \label{fig:geog_dh}
\end{figure}

\begin{figure}[H]
    \centering
    \includegraphics[width=\linewidth]{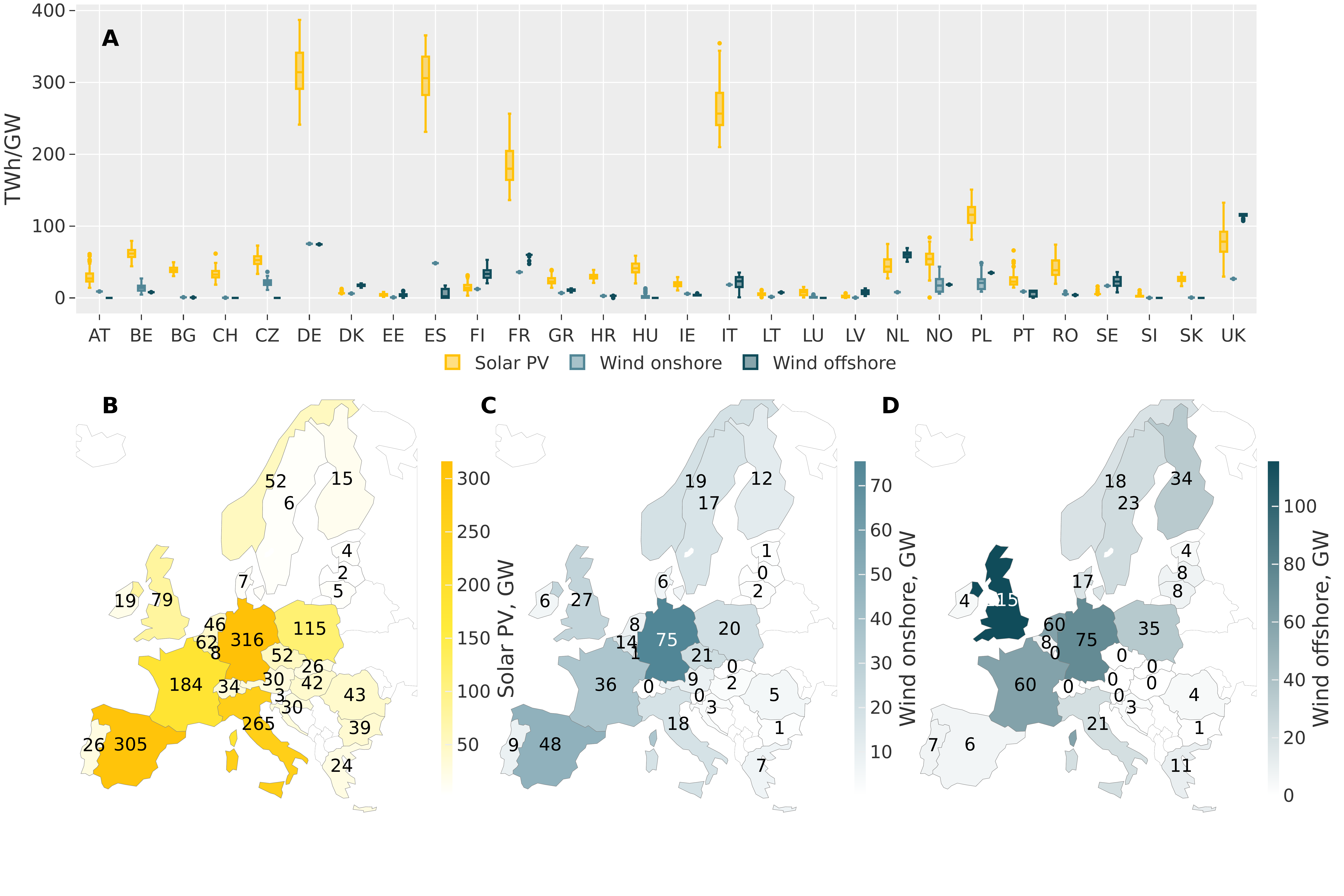}
    \caption{\textit{Geographical distribution of generation capacities - No Heat}: (A) Distribution of solar PV, wind onshore and windoff offshore by country across weather years (1940/41 - 2017/18), (B) - (D) Mean capacities }
    \label{fig:geog_generation_no_heat}
\end{figure}

\begin{figure}[H]
    \centering
    \includegraphics[width=\linewidth]{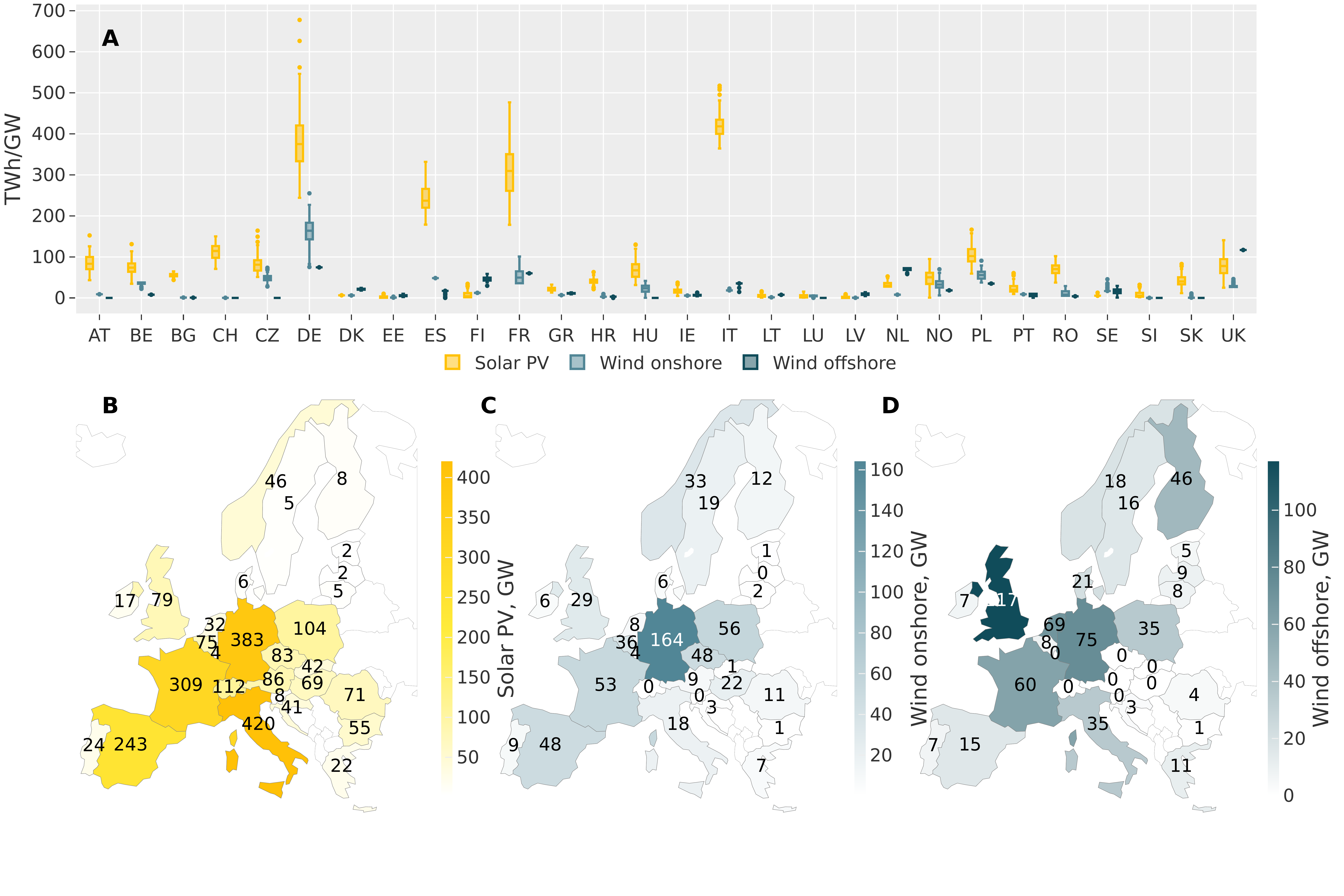}
    \caption{\textit{Geographical distribution of generation capacities - Decent}: (A) Distribution of solar PV, wind onshore and windoff offshore by country across weather years (1940/41 - 2017/18), (B) - (D) Mean capacities }
    \label{fig:geog_generation_decentral}
\end{figure}

\begin{figure}[H]
    \centering
    \includegraphics[width=\linewidth]{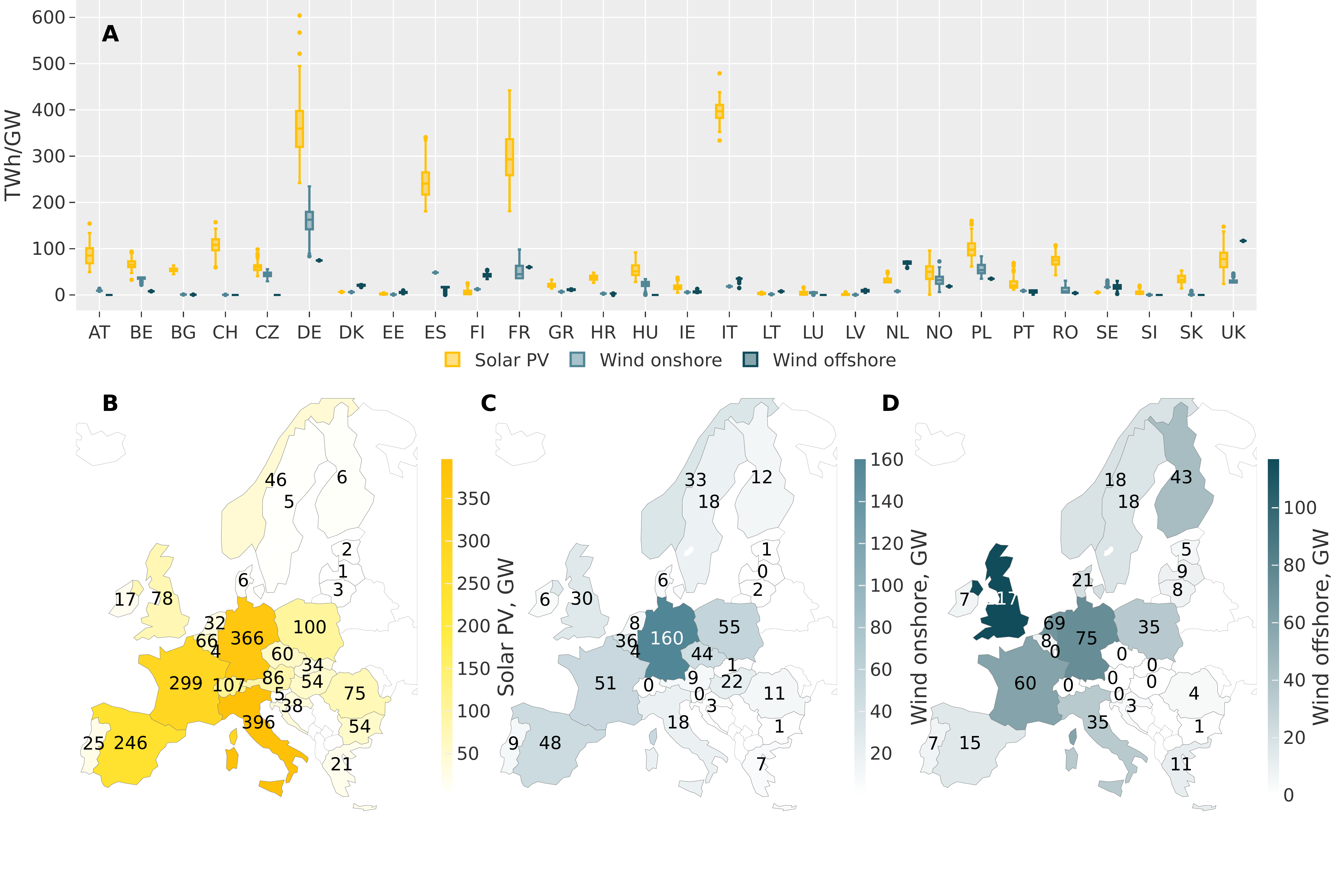}
    \caption{\textit{Geographical distribution of generation capacities - District \& Decent}: (A) Distribution of solar PV, wind onshore and windoff offshore by country across weather years (1940/41 - 2017/18), (B) - (D) Mean capacities }
    \label{fig:geog_generation_dh
    }
\end{figure}

\subsection{Input data}

\subsubsection{Networks, demands and potentials}

\begin{figure}[H]
    \centering
    \includegraphics[width=\linewidth]{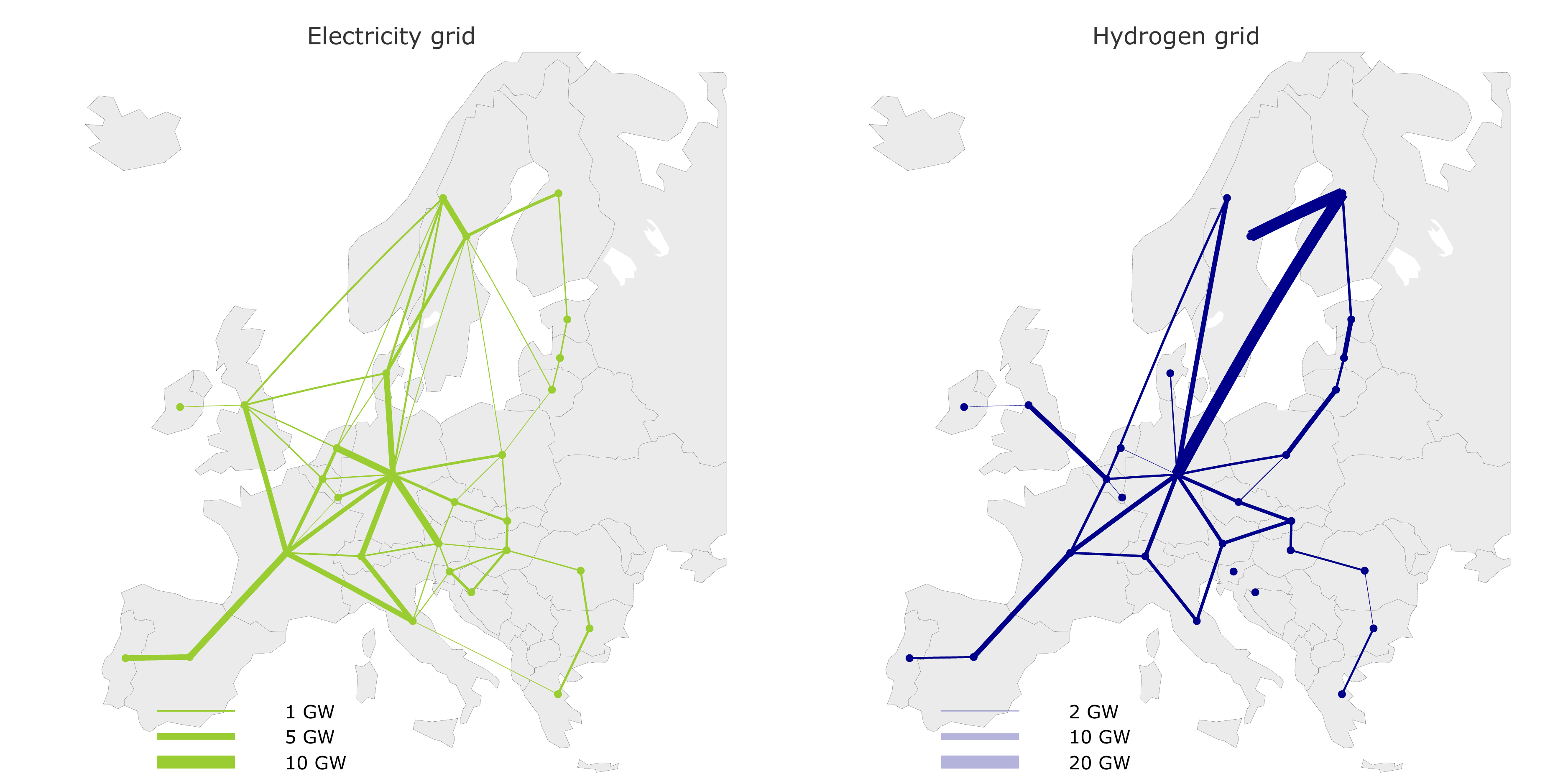}
    \caption{\textit{Assumed network capacities}: \\ \scriptsize{Net transfer capacities in the model are fixed to TYNDP 2022  levels for the electricity grid \cite{entso-e_tyndp_2022} and to TYNDP 2024 levels  for the hydrogen network \cite{entso-e_tyndp_2025}.}} 
    \label{fig:networks}
\end{figure}

\begin{figure}[H]
    \centering
    \includegraphics[width=\linewidth]{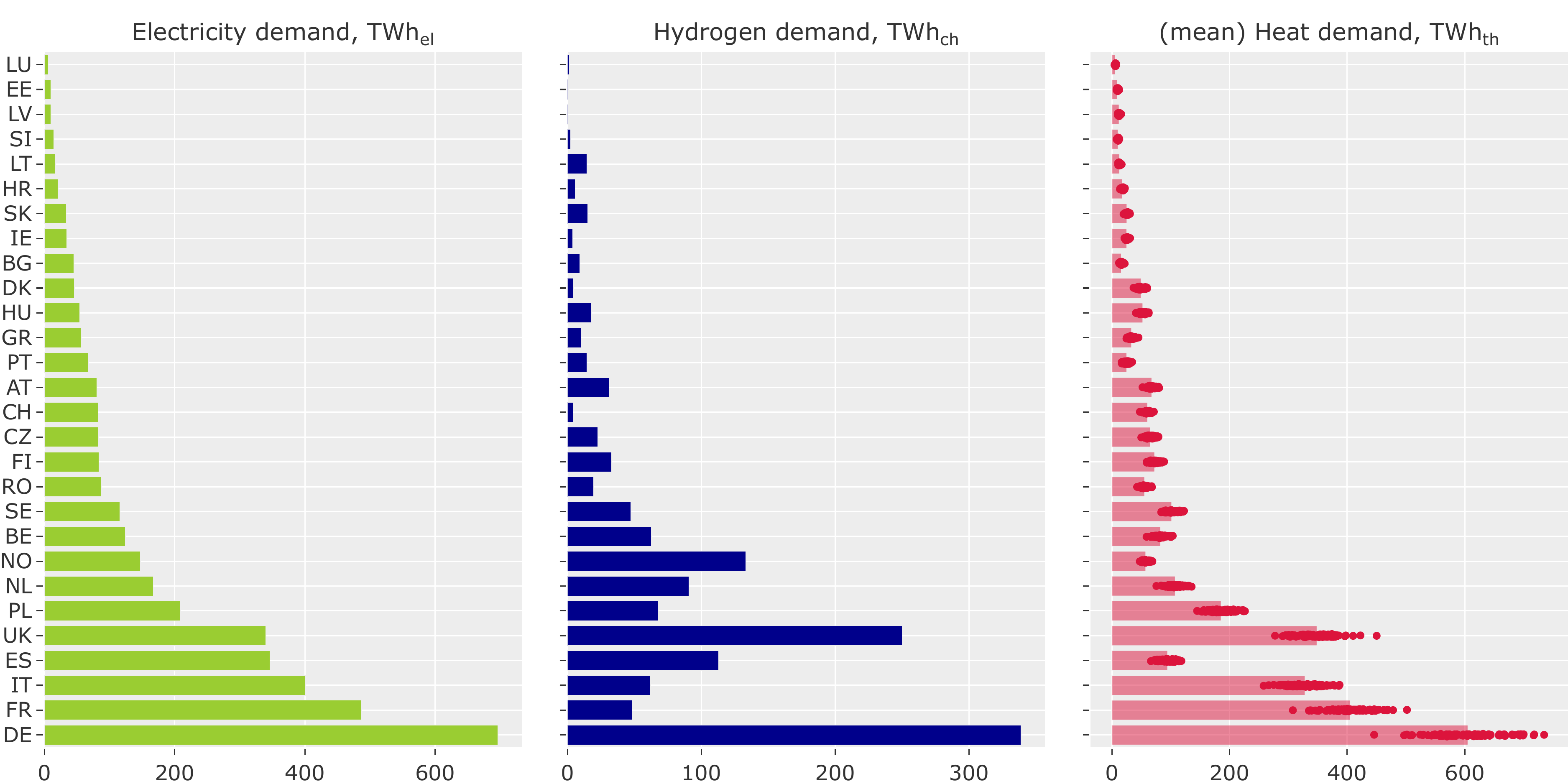}
    \caption{\textit{Annual energy demands by sector}: \\ \scriptsize (left panel) Exogenous electricity demand consisting of baseline electricity demand based on 2017 historical data from Open Power System Data \cite{muehlenpfordt_open_2020}, corrected for existing electrified heat demand using data When2Heat data \cite{ruhnau_update_2022}. (
    (Middle panel): Exogenous hydrogen demand from industry based on TYNDP 2024 \cite{entso-e_tyndp_2025}. (Right panel) Mean aggregate heat demands and year-by year variations based on dataset by Antonioni et al. \cite{antonini_weather-_2024} and the Joint Research Council IDEES database \cite{rozsai_m_jrc-idees-2021_2024}.}
    \label{fig:energy_demand}
\end{figure}

\begin{figure}[H]
    \centering
    \includegraphics[width=\linewidth]{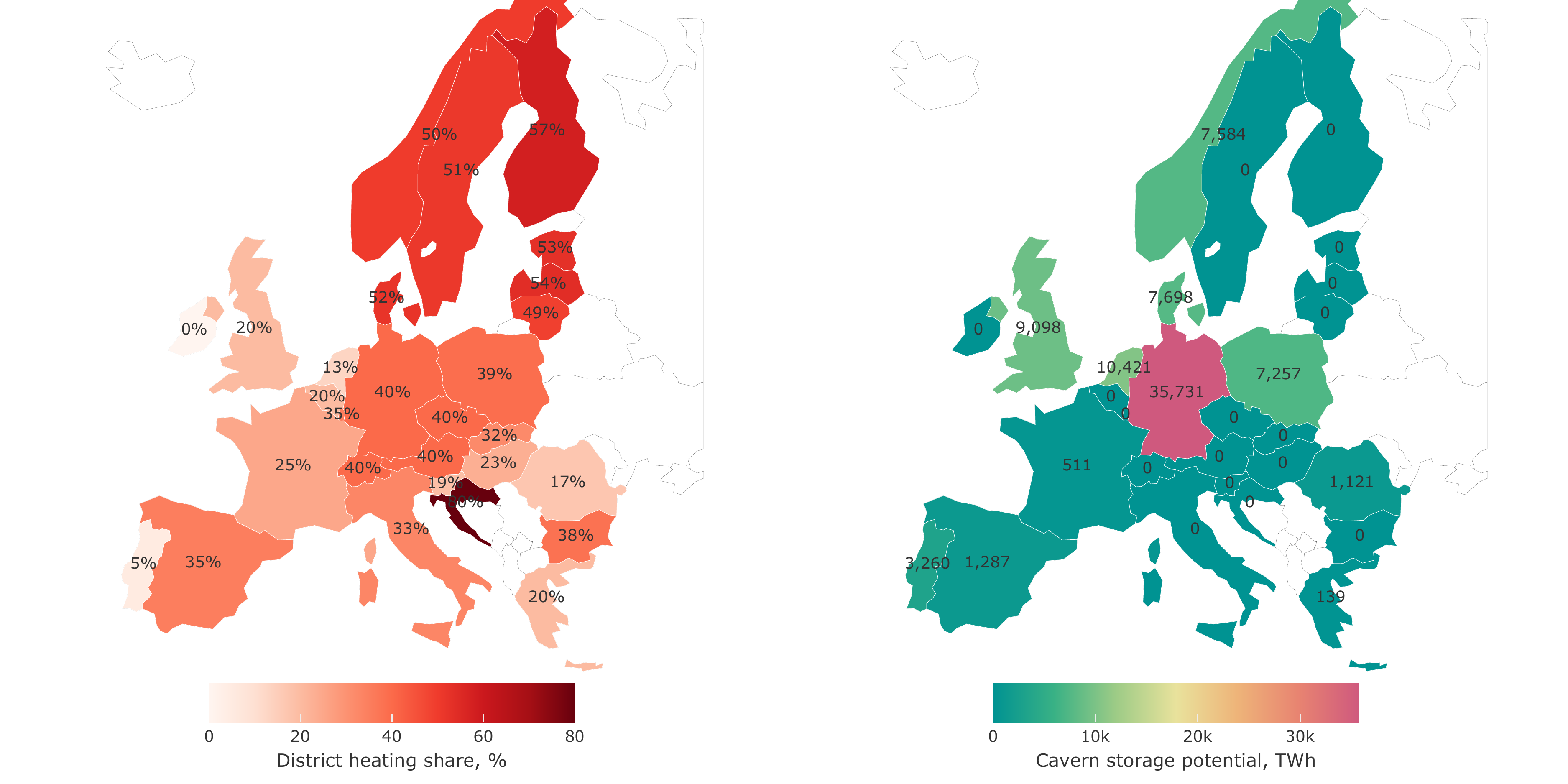}
    \caption{\textit{District heating and cavern storage potentials}:\\
    \scriptsize\textbf{Left panel}: Projected district heating potential by country as share of total heat demand, from \cite{fallahnejad_district_2024}; 
    \textbf{Right panel}: Cavern storage potentials in chemical terawatt hours, from \cite{caglayan_technical_2020}.}
    \label{fig:potentials}
\end{figure}

\subsubsection{Weather data}

We use an open-source dataset of 78 weather years with country-level renewable energy capacity factors for solar PV, wind onshore and wind offshore \cite{antonini_weather-_2024}. The dataset also contains hydro inflows for run-of-river dams and basins. Finally, it also provides temperature-dependent space and water heating profiles for European countries. 

\begin{figure}[H]
    \centering
    \includegraphics[width=\linewidth]{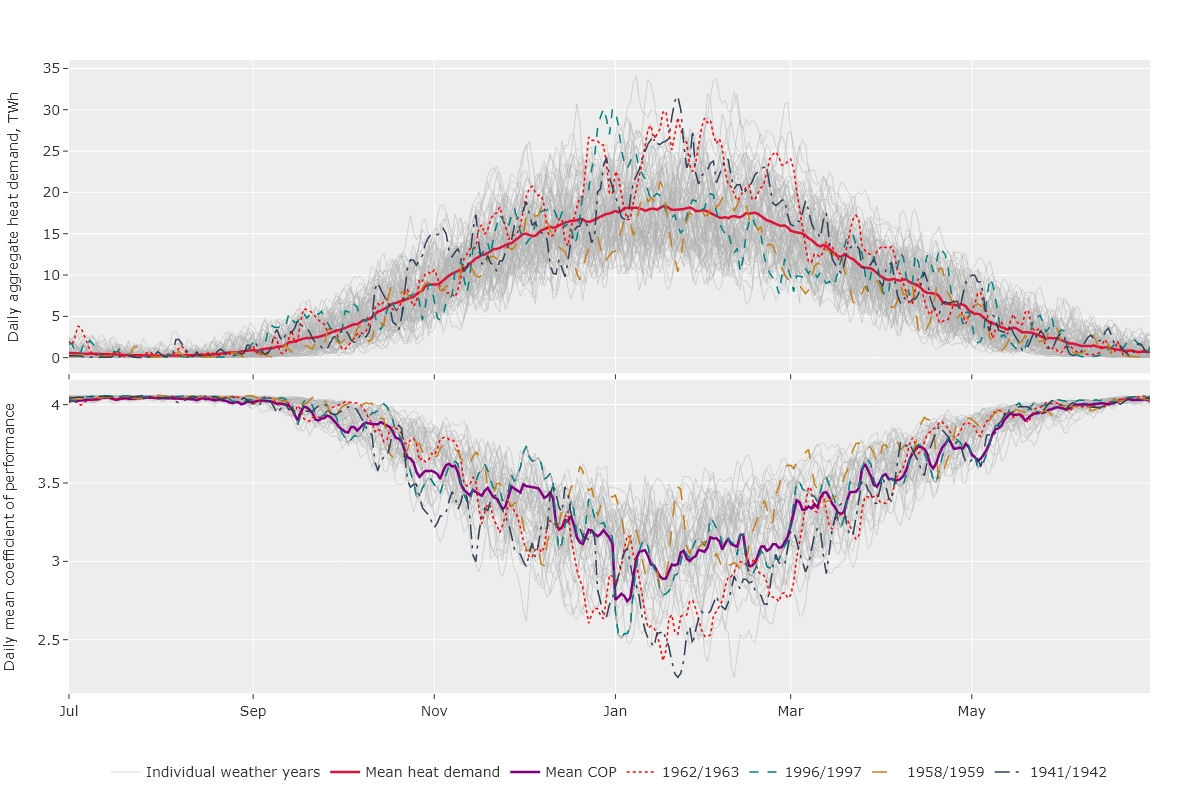}
    \caption{\textit{Heat demand and coefficients of performance}}
    \label{fig:input_heat}
\end{figure}

The dataset does not contain coefficients of performance  for heat pumps. The meteorological data underlying the dataset are from the Copernicus Project's ERA-5 reanalysis database \cite{hersback_era5_2023}. We use the same database to download hourly near-ground temperatures at a granular spatial level. We compute and aggregate coefficients of performance for air-sourced heat pumps to country-level using the When2Heat tool \cite{ruhnau_update_2022} (Figure \ref{fig:input_heat}, lower panel). 

We scale the heat demand profiles by annual country-level heat demands from 2015 as per the JRC-IDEES database \cite{rozsai_m_jrc-idees-2021_2024}. The resulting annual data exhibits a trend, which may reflect both climate change and increased energy efficiency. As we are interested in interannual variations, we estimate a linear model per country to de-trend the data around the 2015 anchor year. The resulting heat demand time series are shown in Figure \ref{fig:input_heat}, upper panel.

For hydro inflow time series, we following Goetske et al. in correcting inflows around a historical break point \cite{gotske_designing_2024}, which we assume to be 1980. Anything prior to that break point will be scaled by the ratio of means of annual inflows of the time series before and after the break point. 

Finally, we pro-rate the hydro inflow time series by capacity to distribute inflows between open pumped-hydro plants and pure reservoirs in case both exist in a country.

\subsection{Technology data}

\begin{table}[H]
    \centering
        \resizebox{\linewidth}{!}{\input{tables/capacity_bounds}}
    \caption{\textit{Capacity bounds of generation capacities}}
    \label{tab:capa_bounds}
\end{table}

\begin{table}[H]
    \centering
        \resizebox{\linewidth}{!}{\input{tables/hydro_capas}}
    \caption{\textit{Capacities of hydro plants}}
    \label{tab:hydro_capas}
\end{table}

\begin{table}[H]
    \centering
        \resizebox{\linewidth}{!}{\input{tables/tech_costs}}
    \caption{\textit{Technology cost assumptions}\\ \scriptsize{ The data is taken from the Technology Catalogues of the Danish Energy Agency \cite{danish_energy_agency_technology_2024}. For pit thermal storage efficiency, the number reflects the efficiency for a 90 day storage duration accounting for heat loss and (dis-)charge losses.}}
    \label{tab:cost_assumptions}
\end{table}

\end{document}

%% file: tables/scenario_table.tex
\begin{tabular}{llll} \toprule
Scenario & Heat demand & LDTS & Description \\ \midrule
     \textit{No Heat} & - & - &  Historical electricity demand corrected for electric heating\\
     \textit{Decent} & Year-specific & - & 80\% of heat demand supplied by decentralized heat pumps \\
     \textit{Decent - Mean} & Long-run mean & - & Like \textit{Decent} but with mean heat demand profile \\
     \textit{District \& Decent} & Year-specific & Yes & 32\% of heat demand supplied by district heating enabling LDTS \\ \bottomrule
\end{tabular}

%% file: tables/sensitivities.tex
\begin{tabular}{llrrrr}
\toprule
\textbf{} & \textbf{} & \textbf{1962/1963} & \textbf{1996/1997} & \textbf{1958/1959} & \textbf{1941/1942} \\
\midrule
\textit{No Heat} & Base case & 40 & 61 & 80 & 27 \\
\midrule
\multirow{4}{*}{\textit{Decent}} 
& Base case & 400 & 341 & 238 & 278 \\
& No H$_2$ network & 433 & 387 & 313 & 315 \\
& Islands & 407 & 363 & 318 & 327 \\
& Nuclear & 359 & 282 & 199 & 247 \\
\midrule
\multirow{5}{*}{\textit{District \& Decent}} 
& Base case & 252 & 237 & 186 & 171 \\
& No H$_2$ network & 301 & 293 & 268 & 221 \\
& Islands & 315 & 300 & 274 & 243 \\
& Nuclear & 219 & 180 & 152 & 144 \\
& All district heating & 137 & 128 & 136 & 98 \\
\bottomrule
\end{tabular}

%% file: tables/capacity_bounds.tex
\begin{tabular}{lrrrrrrrrrrrr}
\toprule
\multirow{2}{*}{Country} 
& \multicolumn{6}{c}{Minimum Capacity (GW)} 
& \multicolumn{6}{c}{Maximum Capacity (GW)} \\
\cline{2-13}
& Biomass & Nuclear & Offshore Wind & Onshore Wind & Run-of-River & Solar PV
& Biomass & Nuclear & Offshore Wind & Onshore Wind & Run-of-River & Solar PV \\
\midrule
AT & 0.0 & 0.0 & 0.0 & 9.0 & 6.1 & 12.0 & 0.6 & 0.0 & 0.0 & 142 & 6.1 & 879 \\
BE & 0.0 & 0.0 & 4.4 & 4.7 & 0.2 & 10.4 & 0.9 & 0.0 & 8.0 & 38 & 0.2 & 499 \\
BG & 0.0 & 2.0 & 0.0 & 1.0 & 0.5 & 3.2 & 0.3 & 2.0 & 0.6 & 254 & 0.5 & 3484 \\
CZ & 0.0 & 4.1 & 0.0 & 1.0 & 0.4 & 3.9 & 0.8 & 4.1 & 0.0 & 203 & 0.4 & 2201 \\
DE & 0.0 & 0.0 & 23.6 & 75.4 & 4.7 & 96.1 & 7.6 & 0.0 & 74.6 & 684 & 4.7 & 8970 \\
DK & 0.0 & 0.0 & 6.8 & 6.2 & 0.0 & 6.5 & 2.3 & 0.0 & 129.2 & 42 & 0.0 & 1269 \\
EE & 0.0 & 0.0 & 0.2 & 0.8 & 0.0 & 0.4 & 0.4 & 0.0 & 10.0 & 180 & 0.0 & 967 \\
ES & 0.0 & 3.0 & 0.2 & 48.4 & 3.6 & 38.4 & 1.7 & 3.0 & 17.4 & 1032 & 3.6 & 18143 \\
FI & 0.0 & 4.4 & 1.5 & 12.4 & 0.0 & 1.5 & 3.7 & 4.4 & 65.0 & 1714 & 0.0 & 5191 \\
FR & 0.0 & 58.2 & 5.5 & 35.9 & 13.6 & 43.4 & 2.4 & 58.2 & 60.0 & 734 & 13.6 & 15942 \\
GR & 0.0 & 0.0 & 0.1 & 6.9 & 0.4 & 7.5 & 0.5 & 0.0 & 22.2 & 314 & 0.4 & 2214 \\
HR & 0.0 & 0.0 & 0.0 & 2.8 & 0.4 & 0.4 & 0.4 & 0.0 & 3.0 & 130 & 0.4 & 1346 \\
HU & 0.0 & 3.1 & 0.0 & 0.3 & 0.1 & 6.5 & 0.9 & 3.1 & 0.0 & 88 & 0.1 & 4701 \\
IE & 0.0 & 0.0 & 3.5 & 5.8 & 0.2 & 0.6 & 0.1 & 0.0 & 35.0 & 111 & 0.2 & 2330 \\
IT & 0.0 & 0.0 & 0.9 & 18.4 & 6.3 & 51.1 & 4.7 & 0.0 & 35.4 & 499 & 6.3 & 4942 \\
LT & 0.0 & 0.0 & 0.7 & 1.5 & 0.1 & 0.9 & 0.2 & 0.0 & 7.6 & 168 & 0.1 & 1911 \\
LU & 0.0 & 0.0 & 0.0 & 0.4 & 0.0 & 0.6 & 0.1 & 0.0 & 0.0 & 6 & 0.0 & 59 \\
LV & 0.0 & 0.0 & 0.5 & 0.4 & 1.6 & 0.0 & 0.2 & 0.0 & 14.0 & 251 & 1.6 & 1482 \\
NL & 0.0 & 0.5 & 11.5 & 8.0 & 0.0 & 27.3 & 3.9 & 0.5 & 72.5 & 25 & 0.0 & 722 \\
NO & 0.0 & 0.0 & 0.2 & 6.1 & 0.0 & 0.6 & 0.0 & 0.0 & 18.5 & 1105 & 0.0 & 8494 \\
PL & 0.0 & 0.0 & 5.9 & 8.7 & 0.5 & 5.1 & 1.5 & 0.0 & 35.0 & 732 & 0.5 & 9949 \\
PT & 0.0 & 0.0 & 0.3 & 8.9 & 0.8 & 7.5 & 1.1 & 0.0 & 10.0 & 142 & 0.8 & 2401 \\
RO & 0.0 & 2.0 & 0.0 & 5.3 & 3.4 & 5.1 & 0.1 & 2.0 & 4.0 & 465 & 3.4 & 8505 \\
SE & 0.0 & 5.9 & 1.0 & 16.9 & 0.0 & 5.4 & 4.6 & 5.9 & 72.0 & 2356 & 0.0 & 7294 \\
SI & 0.0 & 0.7 & 0.0 & 0.2 & 1.3 & 1.9 & 0.1 & 0.7 & 0.0 & 52 & 1.3 & 127 \\
SK & 0.0 & 2.7 & 0.0 & 0.5 & 1.6 & 1.2 & 0.6 & 2.7 & 0.0 & 136 & 1.6 & 1202 \\
CH & 0.0 & 1.2 & 0.0 & 0.3 & 4.2 & 9.8 & 0.4 & 1.2 & 0.0 & 43 & 4.2 & 178 \\
UK & 0.0 & 9.3 & 34.8 & 26.6 & 2.1 & 23.4 & 6.4 & 9.3 & 116.9 & 542 & 2.1 & 6612 \\
\bottomrule
\end{tabular}

%% file: tables/hydro_capas.tex
\begin{tabular}{lrrrrrrrr}
\toprule
\multirow{2}{*}{Country} 
& \multicolumn{2}{c}{Power in (GW)} 
& \multicolumn{3}{c}{Power out (GW)} 
& \multicolumn{3}{c}{Energy (GWh)} \\
\cline{2-9}
& Pumped hydro closed & Pumped hydro open 
& Pumped hydro closed & Pumped hydro open & Reservoir 
& Pumped hydro closed & Pumped hydro open & Reservoir \\
\midrule
AT & 0.30 & 5.21 & 0.30 & 6.04 & 2.49 & 2 & 1733 & 757 \\
BE & 1.15 & & 1.22 & & & 5 & & \\
BG & 0.78 & 0.15 & 0.86 & 0.54 & 1.28 & 9 & 255 & 843 \\
CH & 1.90 & 2.05 & 1.90 & 10.73 & & 70 & 8800 & \\
CZ & 0.66 & 0.44 & 0.69 & 0.47 & 0.70 & 4 & 2 & 3 \\
DE & 7.42 & 1.36 & 7.38 & 1.64 & 1.30 & 242 & 417 & 259 \\
DK & & & & & & & & \\
EE & & & & & & & & \\
ES & 6.68 & 2.42 & 6.87 & 2.68 & 10.97 & 99 & 5962 & 11649 \\
FI & & & & & 3.24 & & & 5530 \\
FR & 1.95 & 1.85 & 1.95 & 1.85 & 8.79 & 10 & 90 & 10000 \\
GR & 0.71 & 0.74 & 0.68 & 0.70 & 2.74 & & 5 & 3600 \\
HR & & 0.25 & & 0.28 & 1.94 & & 18 & 2046 \\
HU & & & & & & & & \\
IE & 0.29 & & 0.29 & & & 2 & & \\
IT & 7.37 & 2.10 & 7.30 & 3.33 & 9.65 & 70 & 309 & 5649 \\
LT & 0.90 & & 0.93 & & & 11 & & \\
LU & 1.03 & & 1.31 & & & 5 & & \\
LV & & & & & & & & \\
NL & & & & & & & & \\
NO & & 1.09 & & & 38.05 & & & 89577 \\
PL & 1.49 & 0.17 & 1.32 & 0.22 & 0.27 & 6 & 2 & 1 \\
PT & & 3.58 & & 3.83 & 3.77 & & 2018 & 1290 \\
RO & & 0.10 & & 0.84 & 2.58 & & 1506 & 2425 \\
SE & & & & & 16.45 & & & 31931 \\
SI & 0.19 & & 0.18 & & & 3 & & \\
SK & 0.65 & 0.18 & 0.65 & 0.28 & 0.02 & 4 & 46 & 17 \\
UK & 2.68 & & 2.95 & & & 26 & & \\
\bottomrule
\end{tabular}

%% file: tables/tech_costs.tex
\begin{tabular}{lrrrrrr}
\toprule
\multirow{2}{*}{Technology} 
& \multicolumn{3}{c}{Costs} 
& \multirow{2}{*}{Efficiency} 
& \multirow{2}{*}{Lifetime} 
& \multirow{2}{*}{Electricity demand} \\
\cline{2-4}
& Investment & Fixed O\&M & Variable+Fuel \\
& [€/kW or €/kWh] & [€/kW/a or €/kWh/a] & [€/MWh] & [\%]  & [years] & [\% of input]\\
\midrule
Biomass & 2517 & 129.01 & 13.65 & 48.7\% & 30 & \\
Nuclear & 8080 & 102.62 & 5.24 & 33.7\% & 40 & \\
H$_2$ turbine & 501 & 7.893 & $p_{H2}$ + 5.00 & 43.0\% & 25 & \\
Offshore wind & 1843 & 37.08 & 3.37 &  & 30 & \\
Onshore wind & 1225 & 17.53 & 2.08 &  & 30 & \\
Run-of-river & 3870 & 77.41 &  &  & 50 & \\
Solar PV & 458 & 8.76 &  &  & 40 & \\
\midrule
Battery inverter & 72 & 0.65 & 2.00 & 96.0\% & 30 & \\
Battery storage & 112 &  &  & 99.9\% & 30 & \\
\midrule
PEM electrolysis & 404 & 8.05 &  & 66.4\% & 20/25 & 0.75\% \\
Hydrogen cavern storage (kWh$_{h2}$) & 1.43 &  &  & 100\% & 100 & \\
H$_2$ cavern compressor (kW$_{h2}$) & 96 & 3.83 &  &  & 15 & 2.10\% \\
Hydrogen tank storage (kWh$_{h2}$) & 17 & 0.48 &  &  & 30 & \\
H$_2$ tank compressor (kW$_{h2}$) & 10 & 0.48 &  &  & 25 & 6.00\% \\
\midrule
Large-scale air-sourced heat pump (kW$_{th}$) & 970 & 2.55 &  & COP & 25 & \\
Pit thermal storage (kWh$_{th}$)& 0.57 & 0.0037 &  & 49.12\%* & 25 & \\
\bottomrule
\end{tabular}

%% file: references.bib
@misc{schmidt_long-duration_2025,
	title = {On long-duration storage, weather uncertainty and limited foresight},
	url = {http://arxiv.org/abs/2505.12538},
	doi = {10.48550/arXiv.2505.12538},
	urldate = {2025-05-20},
	publisher = {arXiv},
	author = {Schmidt, Felix},
	month = may,
	year = {2025},
	note = {arXiv:2505.12538 [econ]},
}

@article{lund_review_2015,
	title = {Review of energy system flexibility measures to enable high levels of variable renewable electricity},
	volume = {45},
	issn = {13640321},
	url = {https://linkinghub.elsevier.com/retrieve/pii/S1364032115000672},
	doi = {10.1016/j.rser.2015.01.057},
	language = {en},
	urldate = {2025-05-19},
	journal = {Renewable and Sustainable Energy Reviews},
	author = {Lund, Peter D. and Lindgren, Juuso and Mikkola, Jani and Salpakari, Jyri},
	month = may,
	year = {2015},
	pages = {785--807},
}

@article{mockert_meteorological_2023,
	title = {Meteorological conditions during periods of low wind speed and insolation in {Germany}: {The} role of weather regimes},
	volume = {30},
	copyright = {© 2023 The Authors. Meteorological Applications published by John Wiley \& Sons Ltd on behalf of Royal Meteorological Society.},
	issn = {1469-8080},
	shorttitle = {Meteorological conditions during periods of low wind speed and insolation in {Germany}},
	doi = {10.1002/met.2141},
	language = {en},
	number = {4},
	urldate = {2025-04-08},
	journal = {Meteorological Applications},
	author = {Mockert, Fabian and Grams, Christian M. and Brown, Tom and Neumann, Fabian},
	year = {2023},
	pages = {e2141},
}

@article{rosenow_meta-review_2024,
	title = {A meta-review of 54 studies on hydrogen heating},
	volume = {1},
	issn = {2949-7906},
	url = {https://www.sciencedirect.com/science/article/pii/S2949790623000101},
	doi = {10.1016/j.crsus.2023.100010},
	number = {1},
	urldate = {2025-05-18},
	journal = {Cell Reports Sustainability},
	author = {Rosenow, Jan},
	month = jan,
	year = {2024},
	pages = {100010},
}

@techreport{luderer_energiewende_2025,
	title = {Die {Energiewende} kosteneffizient gestalten: {Szenarien} zur {Klimaneutralität} 2045},
	shorttitle = {Die {Energiewende} kosteneffizient gestalten},
	url = {https://publications.pik-potsdam.de/pubman/item/item_32090},
	language = {de},
	urldate = {2025-05-18},
	institution = {Potsdam Institute for Climate Impact Research},
	author = {Luderer, Gunnar and Bartels, Frederike and Brown, Tom and Aulich, Clara and Benke, Falk and Fleiter, Tobias and Frank, Fabio and Ganal, Helen and Geis, Julian and Gerhardt, Norman and Gnann, Till and Gunnemann, Alyssa and Hasse, Robin and Herbst, Andrea and Herkel, Sebastian and Hoppe, Johanna and Kost, Christoph and Krail, Michael and Lindner, Michael and Neuwirth, Marius and Nolte, Hannah and Pietzcker, Robert and Plötz, Patrick and Rehfeldt, Matthias and Schreyer, Felix and Seibold, Toni and Senkpiel, Charlotte and Sörgel, Dominika and Speth, Daniel and Steffen, Bjarne and Verpoort, Philipp C.},
	month = mar,
	year = {2025},
	doi = {10.48485/PIK.2025.003},
	note = {Artwork Size: 106 pages, 33,7 MB
Medium: application/pdf},
	pages = {106 pages, 33,7 MB},
}

@article{xiang_comprehensive_2022,
	title = {A comprehensive review on pit thermal energy storage: {Technical} elements, numerical approaches and recent applications},
	volume = {55},
	issn = {2352-152X},
	shorttitle = {A comprehensive review on pit thermal energy storage},
	url = {https://www.sciencedirect.com/science/article/pii/S2352152X22017042},
	doi = {10.1016/j.est.2022.105716},
	urldate = {2025-05-18},
	journal = {Journal of Energy Storage},
	author = {Xiang, Yutong and Xie, Zichan and Furbo, Simon and Wang, Dengjia and Gao, Meng and Fan, Jianhua},
	month = nov,
	year = {2022},
	pages = {105716},
}

@book{tsiropoulos_towards_2020,
	title = {Towards net-zero emissions in the {EU} energy system by 2050 – {Insights} from scenarios in line with the 2030 and 2050 ambitions of the {European} {Green} {Deal}},
	publisher = {Publications Office},
	author = {Tsiropoulos, I. and Nijs, W. and Tarvydas, D. and Ruiz, P.},
	year = {2020},
	doi = {doi/10.2760/081488},
}

@misc{ember_yearly_2024,
	title = {Yearly {Electricity} {Data}},
	url = {https://ember-energy.org/data/yearly-electricity-data},
	language = {en-US},
	urldate = {2025-05-18},
	journal = {Ember},
	author = {{Ember}},
	year = {2024},
}

@article{entso-e_tyndp_2022,
	title = {{TYNDP} 2022 {Scenario} {Building} {Guidelines}},
	language = {en},
	author = {{ENTSO-E}},
	month = apr,
	year = {2022},
}

@article{ostergaard_optimal_2023,
	title = {Optimal heat storage in district energy plants with heat pumps and electrolysers},
	volume = {275},
	issn = {0360-5442},
	url = {https://www.sciencedirect.com/science/article/pii/S0360544223008174},
	doi = {10.1016/j.energy.2023.127423},
	urldate = {2025-05-17},
	journal = {Energy},
	author = {Østergaard, Poul Alberg and Andersen, Anders N.},
	month = jul,
	year = {2023},
	pages = {127423},
}

@article{sundarrajan_harnessing_2025,
	title = {Harnessing hydrogen and thermal energy storage: {Sweden}'s path to a 100 \% renewable energy system by 2045},
	volume = {210},
	issn = {1364-0321},
	shorttitle = {Harnessing hydrogen and thermal energy storage},
	url = {https://www.sciencedirect.com/science/article/pii/S1364032124007676},
	doi = {10.1016/j.rser.2024.115041},
	urldate = {2025-05-17},
	journal = {Renewable and Sustainable Energy Reviews},
	author = {Sundarrajan, Poornima and Thakur, Jagruti and Meha, Drilon},
	month = mar,
	year = {2025},
	pages = {115041},
}

@article{kauko_flexibility_2024,
	title = {Flexibility through power-to-heat in local integrated energy systems with renewable electricity generation and seasonal thermal energy storage},
	volume = {313},
	issn = {0360-5442},
	url = {https://www.sciencedirect.com/science/article/pii/S0360544224037952},
	doi = {10.1016/j.energy.2024.134017},
	urldate = {2025-05-17},
	journal = {Energy},
	author = {Kauko, Hanne and Brækken, August and Askeland, Magnus},
	month = dec,
	year = {2024},
	pages = {134017},
}

@article{victoria_role_2019,
	title = {The role of storage technologies throughout the decarbonisation of the sector-coupled {European} energy system},
	volume = {201},
	issn = {0196-8904},
	url = {https://www.sciencedirect.com/science/article/pii/S0196890419309835},
	doi = {10.1016/j.enconman.2019.111977},
	urldate = {2025-05-17},
	journal = {Energy Conversion and Management},
	author = {Victoria, Marta and Zhu, Kun and Brown, Tom and Andresen, Gorm B. and Greiner, Martin},
	month = dec,
	year = {2019},
	pages = {111977},
}

@article{christensen_role_2024,
	title = {The role of thermal energy storages in future smart energy systems},
	volume = {313},
	issn = {0360-5442},
	url = {https://www.sciencedirect.com/science/article/pii/S0360544224037265},
	doi = {10.1016/j.energy.2024.133948},
	urldate = {2025-05-17},
	journal = {Energy},
	author = {Christensen, Toke Borg Kjær and Lund, Henrik and Sorknæs, Peter},
	month = dec,
	year = {2024},
	pages = {133948},
}

@article{steinbach_future_2024,
	title = {The future {European} hydrogen market: {Market} design and policy recommendations to support market development and commodity trading},
	volume = {70},
	issn = {0360-3199},
	shorttitle = {The future {European} hydrogen market},
	url = {https://www.sciencedirect.com/science/article/pii/S0360319924017944},
	doi = {10.1016/j.ijhydene.2024.05.107},
	urldate = {2025-05-16},
	journal = {International Journal of Hydrogen Energy},
	author = {Steinbach, Sarah A. and Bunk, Nikolas},
	month = jun,
	year = {2024},
	pages = {29--38},
}

@article{odenweller_green_2025,
	title = {The green hydrogen ambition and implementation gap},
	volume = {10},
	copyright = {2025 The Author(s)},
	issn = {2058-7546},
	url = {https://www.nature.com/articles/s41560-024-01684-7},
	doi = {10.1038/s41560-024-01684-7},
	language = {en},
	number = {1},
	urldate = {2025-05-16},
	journal = {Nature Energy},
	author = {Odenweller, Adrian and Ueckerdt, Falko},
	month = jan,
	year = {2025},
	note = {Publisher: Nature Publishing Group},
	pages = {110--123},
}

@article{schill_electricity_2020,
	title = {Electricity {Storage} and the {Renewable} {Energy} {Transition}},
	volume = {4},
	issn = {2542-4351},
	url = {https://www.sciencedirect.com/science/article/pii/S2542435120303408},
	doi = {10.1016/j.joule.2020.07.022},
	number = {10},
	urldate = {2025-05-16},
	journal = {Joule},
	author = {Schill, Wolf-Peter},
	month = oct,
	year = {2020},
	pages = {2059--2064},
}

@article{roth_geographical_2023,
	title = {Geographical balancing of wind power decreases storage needs in a 100\% renewable {European} power sector},
	volume = {26},
	issn = {2589-0042},
	url = {https://www.sciencedirect.com/science/article/pii/S2589004223011513},
	doi = {10.1016/j.isci.2023.107074},
	number = {7},
	urldate = {2025-05-16},
	journal = {iScience},
	author = {Roth, Alexander and Schill, Wolf-Peter},
	month = jul,
	year = {2023},
	pages = {107074},
}

@article{gueret_impacts_2024,
	title = {Impacts of electric carsharing on a power sector with variable renewables},
	volume = {1},
	issn = {2949-7906},
	url = {https://www.sciencedirect.com/science/article/pii/S2949790624003823},
	doi = {10.1016/j.crsus.2024.100241},
	number = {11},
	urldate = {2025-05-16},
	journal = {Cell Reports Sustainability},
	author = {Guéret, Adeline and Schill, Wolf-Peter and Gaete-Morales, Carlos},
	month = nov,
	year = {2024},
	pages = {100241},
}

@techreport{agency_for_the_cooperation_of_energy_regulators_report_2022,
	title = {Report on {Gas} {Storage} {Regulation} and {Indicators}},
	url = {https://www.acer.europa.eu/sites/default/files/documents/Publications/ACER%20Report%20on%20Gas%20Storage%20Regulation%20and%20Indicators.pdf},
	urldate = {2025-05-16},
	author = {{Agency for the Cooperation of Energy Regulators}},
	month = apr,
	year = {2022},
}

@article{oni_underground_2025,
	title = {Underground hydrogen storage in salt caverns: {Recent} advances, modeling approaches, barriers, and future outlook},
	volume = {107},
	issn = {2352-152X},
	shorttitle = {Underground hydrogen storage in salt caverns},
	url = {https://www.sciencedirect.com/science/article/pii/S2352152X24045377},
	doi = {10.1016/j.est.2024.114951},
	urldate = {2025-05-16},
	journal = {Journal of Energy Storage},
	author = {Oni, Babalola Aisosa and Bade, Shree Om and Sanni, Samuel Eshorame and Orodu, Oyinkepreye David},
	month = jan,
	year = {2025},
	pages = {114951},
}

@article{sifnaios_heat_2023,
	title = {Heat losses in water pit thermal energy storage systems in the presence of groundwater},
	volume = {235},
	issn = {1359-4311},
	url = {https://www.sciencedirect.com/science/article/pii/S1359431123014114},
	doi = {10.1016/j.applthermaleng.2023.121382},
	urldate = {2025-05-14},
	journal = {Applied Thermal Engineering},
	author = {Sifnaios, Ioannis and Jensen, Adam R. and Furbo, Simon and Fan, Jianhua},
	month = nov,
	year = {2023},
	pages = {121382},
}

@article{rozsai_m_jrc-idees-2021_2024,
	title = {{JRC}-{IDEES}-2021: the {Integrated} {Database} of the {European} {Energy} {System} – {Data} update and technical documentation},
	issn = {1831-9424 (online)},
	doi = {10.2760/614599 (online)},
	number = {KJ-NA-31-940-EN-N (online)},
	author = {{Rózsai M} and {Jaxa-Rozen M} and {Salvucci R} and {Sikora P} and {Tattini J} and {Neuwahl F}},
	year = {2024},
	note = {ISBN: 978-92-68-16154-8 (online)
Place: Luxembourg (Luxembourg)
Publisher: Publications Office of the European Union},
}

@article{greatbatch_tropical_2015,
	title = {Tropical origin of the severe {European} winter of 1962/1963},
	volume = {141},
	issn = {1477-870X},
	url = {https://onlinelibrary.wiley.com/doi/abs/10.1002/qj.2346},
	doi = {10.1002/qj.2346},
	language = {en},
	number = {686},
	urldate = {2025-05-04},
	journal = {Quarterly Journal of the Royal Meteorological Society},
	author = {Greatbatch, R. J. and Gollan, G. and Jung, T. and Kunz, T.},
	year = {2015},
	note = {\_eprint: https://onlinelibrary.wiley.com/doi/pdf/10.1002/qj.2346},
	pages = {153--165},
}

@article{spatolisano_ammonia_2023,
	title = {Ammonia as a {Carbon}-{Free} {Energy} {Carrier}: {NH3} {Cracking} to {H2}},
	volume = {62},
	issn = {0888-5885},
	shorttitle = {Ammonia as a {Carbon}-{Free} {Energy} {Carrier}},
	url = {https://doi.org/10.1021/acs.iecr.3c01419},
	doi = {10.1021/acs.iecr.3c01419},
	number = {28},
	urldate = {2025-04-09},
	journal = {Industrial \& Engineering Chemistry Research},
	author = {Spatolisano, Elvira and Pellegrini, Laura A. and de Angelis, Alberto R. and Cattaneo, Simone and Roccaro, Ernesto},
	month = jul,
	year = {2023},
	note = {Publisher: American Chemical Society},
	pages = {10813--10827},
}

@misc{hersback_era5_2023,
	title = {{ERA5} hourly data on single levels from 1940 to present},
	url = {https://cds.climate.copernicus.eu/datasets/reanalysis-era5-single-levels?tab=overview},
	doi = {10.24381/cds.adbb2d47},
	language = {en},
	urldate = {2025-04-09},
	publisher = {Copernicus Climate Change Service (C3S) Climate Data Store (CDS)},
	author = {Hersback, H. and Bell, B. and Berrisford, P. and Biavati, G. and Horanyi, A. and Munoz Sabater, J. and Nicolas, J. and Peubey, C. and Radu, R. and Schepers, D. and Simmons, A. and Soci, C. and Dee, D. and Thepaut, J-N.},
	year = {2023},
}

@article{antonini_weather-_2024,
	title = {Weather- and climate-driven power supply and demand time series for power and energy system analyses},
	volume = {11},
	copyright = {2024 The Author(s)},
	issn = {2052-4463},
	url = {https://www.nature.com/articles/s41597-024-04129-8},
	doi = {10.1038/s41597-024-04129-8},
	language = {en},
	number = {1},
	urldate = {2025-04-08},
	journal = {Scientific Data},
	author = {Antonini, Enrico G. A. and Di Bella, Alice and Savelli, Iacopo and Drouet, Laurent and Tavoni, Massimo},
	month = dec,
	year = {2024},
	note = {Publisher: Nature Publishing Group},
	pages = {1324},
}

@misc{danish_energy_agency_technology_2024,
	title = {Technology {Data} for {Generation} of {Electricity} and {District} {Heating} {\textbar} {Energistyrelsen}},
	url = {https://ens.dk/en/analyses-and-statistics/technology-data-generation-electricity-and-district-heating},
	language = {en},
	urldate = {2025-04-08},
	author = {Danish Energy Agency},
	month = dec,
	year = {2024},
}

@article{neumann_potential_2023,
	title = {The potential role of a hydrogen network in {Europe}},
	volume = {7},
	issn = {25424351},
	url = {https://linkinghub.elsevier.com/retrieve/pii/S2542435123002660},
	doi = {10.1016/j.joule.2023.06.016},
	language = {en},
	number = {8},
	urldate = {2025-04-08},
	journal = {Joule},
	author = {Neumann, Fabian and Zeyen, Elisabeth and Victoria, Marta and Brown, Tom},
	month = aug,
	year = {2023},
	pages = {1793--1817},
}

@article{ruhnau_update_2022,
	title = {Update and extension of the {When2Heat} dataset},
	language = {en},
	journal = {ZBW – Leibniz Information Centre for Economics, Kiel, Hamburg},
	author = {Ruhnau, Oliver and Muessel, Jarusch},
	year = {2022},
}

@misc{muehlenpfordt_open_2020,
	title = {Open {Power} {System} {Data}: {Time} series},
	url = {https://data.open-power-system-data.org/time_series/2020-10-06},
	doi = {10.25832/TIME_SERIES/2020-10-06},
	urldate = {2025-04-08},
	publisher = {Open Power System Data},
	author = {Muehlenpfordt, Jonathan},
	month = oct,
	year = {2020},
}

@misc{de_felice_entso-e_2022,
	title = {{ENTSO}-{E} {Pan}-{European} {Climatic} {Database} ({PECD} 2021.3) in {Parquet} format},
	url = {https://zenodo.org/records/7224854},
	doi = {10.5281/zenodo.7224854},
	urldate = {2025-04-08},
	publisher = {Zenodo},
	author = {De Felice, Matteo},
	month = oct,
	year = {2022},
}

@article{trondle_home-made_2019,
	title = {Home-made or imported: {On} the possibility for renewable electricity autarky on all scales in {Europe}},
	volume = {26},
	issn = {2211467X},
	shorttitle = {Home-made or imported},
	url = {https://linkinghub.elsevier.com/retrieve/pii/S2211467X19300811},
	doi = {10.1016/j.esr.2019.100388},
	language = {en},
	urldate = {2025-04-08},
	journal = {Energy Strategy Reviews},
	author = {Tröndle, Tim and Pfenninger, Stefan and Lilliestam, Johan},
	month = nov,
	year = {2019},
	pages = {100388},
}

@article{caglayan_technical_2020,
	title = {Technical potential of salt caverns for hydrogen storage in {Europe}},
	volume = {45},
	issn = {0360-3199},
	url = {https://www.sciencedirect.com/science/article/pii/S0360319919347299},
	doi = {10.1016/j.ijhydene.2019.12.161},
	number = {11},
	urldate = {2025-04-08},
	journal = {International Journal of Hydrogen Energy},
	author = {Caglayan, Dilara Gulcin and Weber, Nikolaus and Heinrichs, Heidi U. and Linßen, Jochen and Robinius, Martin and Kukla, Peter A. and Stolten, Detlef},
	month = feb,
	year = {2020},
	pages = {6793--6805},
}

@article{fallahnejad_district_2024,
	title = {District heating potential in the {EU}-27: {Evaluating} the impacts of heat demand reduction and market share growth},
	volume = {353},
	issn = {0306-2619},
	shorttitle = {District heating potential in the {EU}-27},
	url = {https://www.sciencedirect.com/science/article/pii/S0306261923015180},
	doi = {10.1016/j.apenergy.2023.122154},
	urldate = {2025-04-08},
	journal = {Applied Energy},
	author = {Fallahnejad, Mostafa and Kranzl, Lukas and Haas, Reinhard and Hummel, Marcus and Müller, Andreas and García, Luis Sánchez and Persson, Urban},
	month = jan,
	year = {2024},
	pages = {122154},
}

@article{zerrahn_economics_2018,
	title = {On the economics of electrical storage for variable renewable energy sources},
	volume = {108},
	issn = {0014-2921},
	url = {https://www.sciencedirect.com/science/article/pii/S0014292118301107},
	doi = {10.1016/j.euroecorev.2018.07.004},
	urldate = {2025-04-08},
	journal = {European Economic Review},
	author = {Zerrahn, Alexander and Schill, Wolf-Peter and Kemfert, Claudia},
	month = sep,
	year = {2018},
	pages = {259--279},
}

@techreport{entso-e_tyndp_2025,
	title = {{TYNDP} 2024 // {Scenarios} {Methodology} {Report} – {Final} {Version} {January} 2025},
	language = {en},
	institution = {ENTSO-E},
	author = {ENTSO-E},
	month = jan,
	year = {2025},
}

@article{zerrahn_long-run_2017,
	title = {Long-run power storage requirements for high shares of renewables: review and a new model},
	volume = {79},
	issn = {1364-0321},
	shorttitle = {Long-run power storage requirements for high shares of renewables},
	url = {https://www.sciencedirect.com/science/article/pii/S1364032116308619},
	doi = {10.1016/j.rser.2016.11.098},
	urldate = {2025-04-08},
	journal = {Renewable and Sustainable Energy Reviews},
	author = {Zerrahn, Alexander and Schill, Wolf-Peter},
	month = nov,
	year = {2017},
	pages = {1518--1534},
}

@article{bloess_power--heat_2018,
	title = {Power-to-heat for renewable energy integration: {A} review of technologies, modeling approaches, and flexibility potentials},
	volume = {212},
	issn = {0306-2619},
	shorttitle = {Power-to-heat for renewable energy integration},
	url = {https://www.sciencedirect.com/science/article/pii/S0306261917317889},
	doi = {10.1016/j.apenergy.2017.12.073},
	urldate = {2025-04-08},
	journal = {Applied Energy},
	author = {Bloess, Andreas and Schill, Wolf-Peter and Zerrahn, Alexander},
	month = feb,
	year = {2018},
	pages = {1611--1626},
}

@misc{kittel_quantifying_2024,
	title = {Quantifying the {Dunkelflaute}: {An} analysis of variable renewable energy droughts in {Europe}},
	shorttitle = {Quantifying the {Dunkelflaute}},
	url = {http://arxiv.org/abs/2410.00244},
	doi = {10.48550/arXiv.2410.00244},
	urldate = {2025-04-08},
	publisher = {arXiv},
	author = {Kittel, Martin and Schill, Wolf-Peter},
	month = sep,
	year = {2024},
	note = {arXiv:2410.00244 [eess]},
}

@misc{roth_power_2023,
	title = {Power sector impacts of a simultaneous {European} heat pump rollout},
	url = {http://arxiv.org/abs/2312.06589},
	doi = {10.48550/arXiv.2312.06589},
	urldate = {2025-04-08},
	publisher = {arXiv},
	author = {Roth, Alexander},
	month = dec,
	year = {2023},
	note = {arXiv:2312.06589 [econ]},
}

@article{roth_power_2024,
	title = {Power sector benefits of flexible heat pumps in 2030 scenarios},
	volume = {5},
	copyright = {2024 The Author(s)},
	issn = {2662-4435},
	url = {https://www.nature.com/articles/s43247-024-01861-2},
	doi = {10.1038/s43247-024-01861-2},
	language = {en},
	number = {1},
	urldate = {2025-04-08},
	journal = {Communications Earth \& Environment},
	author = {Roth, Alexander and Gaete-Morales, Carlos and Kirchem, Dana and Schill, Wolf-Peter},
	month = nov,
	year = {2024},
	note = {Publisher: Nature Publishing Group},
	pages = {1--12},
}

@misc{commission_renewable_2023,
	title = {Renewable heating and cooling pathways – {Towards} full decarbonisation by 2050 – {Final} report},
	publisher = {Publications Office of the European Union},
	author = {Commission, European and Energy, Directorate-General for and {E-Think} and ISI, Fraunhofer and University, Halmstad and Wien, T. U. and {Öko-Institut} and Braungardt, S. and Bürger, V. and Fleiter, T. and Bagheri, M. and Manz, P. and Billerbeck, A. and Al-Dabbas, K. and Breitschopf, B. and Winkler, J. and Fallahnejad, M. and Harringer, D. and Hasani, J. and Kök, A. and Kranzl, L. and Mascherbauer, P. and Hummel, M. and Müller, A. and Habiger, J. and Persson, U. and Sánchez-García, L.},
	year = {2023},
	doi = {doi/10.2833/036342},
}

@article{brown_ultra-long-duration_2023,
	title = {Ultra-long-duration energy storage anywhere: {Methanol} with carbon cycling},
	volume = {7},
	issn = {2542-4351},
	shorttitle = {Ultra-long-duration energy storage anywhere},
	url = {https://www.sciencedirect.com/science/article/pii/S2542435123004075},
	doi = {10.1016/j.joule.2023.10.001},
	number = {11},
	urldate = {2025-04-08},
	journal = {Joule},
	author = {Brown, Tom and Hampp, Johannes},
	month = nov,
	year = {2023},
	pages = {2414--2420},
}

@article{sepulveda_design_2021,
	title = {The design space for long-duration energy storage in decarbonized power systems},
	volume = {6},
	copyright = {2021 The Author(s), under exclusive licence to Springer Nature Limited},
	issn = {2058-7546},
	url = {https://www.nature.com/articles/s41560-021-00796-8},
	doi = {10.1038/s41560-021-00796-8},
	language = {en},
	number = {5},
	urldate = {2025-04-08},
	journal = {Nature Energy},
	author = {Sepulveda, Nestor A. and Jenkins, Jesse D. and Edington, Aurora and Mallapragada, Dharik S. and Lester, Richard K.},
	month = may,
	year = {2021},
	note = {Publisher: Nature Publishing Group},
	pages = {506--516},
}

@article{dowling_long-duration_2021,
	title = {Long-duration energy storage for reliable renewable electricity: {The} realistic possibilities},
	volume = {77},
	issn = {0096-3402},
	shorttitle = {Long-duration energy storage for reliable renewable electricity},
	url = {https://doi.org/10.1080/00963402.2021.1989191},
	doi = {10.1080/00963402.2021.1989191},
	number = {6},
	urldate = {2024-10-25},
	journal = {Bulletin of the Atomic Scientists},
	author = {Dowling, Jacqueline A. and Lewis, Nathan S.},
	month = nov,
	year = {2021},
	note = {Publisher: Routledge
\_eprint: https://doi.org/10.1080/00963402.2021.1989191},
	pages = {281--284},
}

@article{jenkins_long-duration_2021,
	title = {Long-duration energy storage: {A} blueprint for research and innovation},
	volume = {5},
	issn = {2542-4785, 2542-4351},
	shorttitle = {Long-duration energy storage},
	url = {https://www.cell.com/joule/abstract/S2542-4351(21)00358-5},
	doi = {10.1016/j.joule.2021.08.002},
	language = {English},
	number = {9},
	urldate = {2024-10-25},
	journal = {Joule},
	author = {Jenkins, Jesse D. and Sepulveda, Nestor A.},
	month = sep,
	year = {2021},
	note = {Publisher: Elsevier},
	pages = {2241--2246},
}

@article{staadecker_value_2024,
	title = {The value of long-duration energy storage under various grid conditions in a zero-emissions future},
	volume = {15},
	copyright = {2024 The Author(s)},
	issn = {2041-1723},
	url = {https://www.nature.com/articles/s41467-024-53274-6},
	doi = {10.1038/s41467-024-53274-6},
	language = {en},
	number = {1},
	urldate = {2024-11-19},
	journal = {Nature Communications},
	author = {Staadecker, Martin and Szinai, Julia and Sánchez-Pérez, Pedro A. and Kurtz, Sarah and Hidalgo-Gonzalez, Patricia},
	month = nov,
	year = {2024},
	note = {Publisher: Nature Publishing Group},
	pages = {9501},
}

@article{ruhnau_storage_2022,
	title = {Storage requirements in a 100\% renewable electricity system: extreme events and inter-annual variability},
	volume = {17},
	issn = {1748-9326},
	shorttitle = {Storage requirements in a 100\% renewable electricity system},
	url = {https://dx.doi.org/10.1088/1748-9326/ac4dc8},
	doi = {10.1088/1748-9326/ac4dc8},
	language = {en},
	number = {4},
	urldate = {2023-10-14},
	journal = {Environmental Research Letters},
	author = {Ruhnau, Oliver and Qvist, Staffan},
	month = mar,
	year = {2022},
	note = {Publisher: IOP Publishing},
	pages = {044018},
}

@article{dowling_role_2020,
	title = {Role of {Long}-{Duration} {Energy} {Storage} in {Variable} {Renewable} {Electricity} {Systems}},
	volume = {4},
	issn = {25424351},
	url = {https://linkinghub.elsevier.com/retrieve/pii/S2542435120303251},
	doi = {10.1016/j.joule.2020.07.007},
	language = {en},
	number = {9},
	urldate = {2023-09-26},
	journal = {Joule},
	author = {Dowling, Jacqueline A. and Rinaldi, Katherine Z. and Ruggles, Tyler H. and Davis, Steven J. and Yuan, Mengyao and Tong, Fan and Lewis, Nathan S. and Caldeira, Ken},
	month = sep,
	year = {2020},
	pages = {1907--1928},
}

@article{brown_synergies_2018,
	title = {Synergies of sector coupling and transmission reinforcement in a cost-optimised, highly renewable {European} energy system},
	volume = {160},
	issn = {0360-5442},
	url = {https://www.sciencedirect.com/science/article/pii/S036054421831288X},
	doi = {10.1016/j.energy.2018.06.222},
	urldate = {2025-04-08},
	journal = {Energy},
	author = {Brown, T. and Schlachtberger, D. and Kies, A. and Schramm, S. and Greiner, M.},
	month = oct,
	year = {2018},
	pages = {720--739},
}

@article{alizadeh_flexibility_2016,
	title = {Flexibility in future power systems with high renewable penetration: {A} review},
	volume = {57},
	issn = {1364-0321},
	shorttitle = {Flexibility in future power systems with high renewable penetration},
	url = {https://www.sciencedirect.com/science/article/pii/S136403211501583X},
	doi = {10.1016/j.rser.2015.12.200},
	urldate = {2025-04-08},
	journal = {Renewable and Sustainable Energy Reviews},
	author = {Alizadeh, M. I. and Parsa Moghaddam, M. and Amjady, N. and Siano, P. and Sheikh-El-Eslami, M. K.},
	month = may,
	year = {2016},
	pages = {1186--1193},
}

@misc{kittel_coping_2025,
	title = {Coping with the {Dunkelflaute}: {Power} system implications of variable renewable energy droughts in {Europe}},
	shorttitle = {Coping with the {Dunkelflaute}},
	url = {http://arxiv.org/abs/2411.17683},
	doi = {10.48550/arXiv.2411.17683},
	urldate = {2025-04-08},
	publisher = {arXiv},
	author = {Kittel, Martin and Roth, Alexander and Schill, Wolf-Peter},
	month = jan,
	year = {2025},
	note = {arXiv:2411.17683 [physics]},
}

@article{kittel_measuring_2024,
	title = {Measuring the {Dunkelflaute}: how (not) to analyze variable renewable energy shortage},
	volume = {1},
	issn = {2753-3751},
	shorttitle = {Measuring the {Dunkelflaute}},
	url = {https://dx.doi.org/10.1088/2753-3751/ad6dfc},
	doi = {10.1088/2753-3751/ad6dfc},
	language = {en},
	number = {3},
	urldate = {2025-04-08},
	journal = {Environmental Research: Energy},
	author = {Kittel, Martin and Schill, Wolf-Peter},
	month = aug,
	year = {2024},
	note = {Publisher: IOP Publishing},
	pages = {035007},
}

@article{craig_overcoming_2022,
	title = {Overcoming the disconnect between energy system and climate modeling},
	volume = {6},
	issn = {2542-4785, 2542-4351},
	url = {https://www.cell.com/joule/abstract/S2542-4351(22)00237-9},
	doi = {10.1016/j.joule.2022.05.010},
	language = {English},
	number = {7},
	urldate = {2025-04-08},
	journal = {Joule},
	author = {Craig, Michael T. and Wohland, Jan and Stoop, Laurens P. and Kies, Alexander and Pickering, Bryn and Bloomfield, Hannah C. and Browell, Jethro and Felice, Matteo De and Dent, Chris J. and Deroubaix, Adrien and Frischmuth, Felix and Gonzalez, Paula L. M. and Grochowicz, Aleksander and Gruber, Katharina and Härtel, Philipp and Kittel, Martin and Kotzur, Leander and Labuhn, Inga and Lundquist, Julie K. and Pflugradt, Noah and Wiel, Karin van der and Zeyringer, Marianne and Brayshaw, David J.},
	month = jul,
	year = {2022},
	note = {Publisher: Elsevier},
	pages = {1405--1417},
}

@article{ruggles_planning_2024,
	title = {Planning reliable wind- and solar-based electricity systems},
	volume = {15},
	issn = {2666-7924},
	url = {https://www.sciencedirect.com/science/article/pii/S2666792424000234},
	doi = {10.1016/j.adapen.2024.100185},
	urldate = {2025-04-08},
	journal = {Advances in Applied Energy},
	author = {Ruggles, Tyler H. and Virgüez, Edgar and Reich, Natasha and Dowling, Jacqueline and Bloomfield, Hannah and Antonini, Enrico G. A. and Davis, Steven J. and Lewis, Nathan S. and Caldeira, Ken},
	month = sep,
	year = {2024},
	pages = {100185},
}

@article{goke_stabilized_2024,
	title = {Stabilized {Benders} decomposition for energy planning under climate uncertainty},
	volume = {316},
	issn = {0377-2217},
	url = {https://www.sciencedirect.com/science/article/pii/S0377221724000353},
	doi = {10.1016/j.ejor.2024.01.016},
	number = {1},
	urldate = {2025-04-08},
	journal = {European Journal of Operational Research},
	author = {Göke, Leonard and Schmidt, Felix and Kendziorski, Mario},
	month = jul,
	year = {2024},
	pages = {183--199},
}

@article{grochowicz_intersecting_2023,
	title = {Intersecting near-optimal spaces: {European} power systems with more resilience to weather variability},
	volume = {118},
	issn = {0140-9883},
	shorttitle = {Intersecting near-optimal spaces},
	url = {https://www.sciencedirect.com/science/article/pii/S0140988322006259},
	doi = {10.1016/j.eneco.2022.106496},
	urldate = {2025-04-08},
	journal = {Energy Economics},
	author = {Grochowicz, Aleksander and van Greevenbroek, Koen and Benth, Fred Espen and Zeyringer, Marianne},
	month = feb,
	year = {2023},
	pages = {106496},
}

@article{bloomfield_quantifying_2016,
	title = {Quantifying the increasing sensitivity of power systems to climate variability},
	volume = {11},
	url = {https://iopscience.iop.org/article/10.1088/1748-9326/11/12/124025/meta},
	number = {12},
	urldate = {2025-04-08},
	journal = {Environmental Research Letters},
	author = {Bloomfield, Hannah C. and Brayshaw, David J. and Shaffrey, Len C. and Coker, Phil J. and Thornton, Hazel E.},
	year = {2016},
	note = {Publisher: IOP Publishing},
	pages = {124025},
}

@article{van_der_wiel_influence_2019,
	title = {The influence of weather regimes on {European} renewable energy production and demand},
	volume = {14},
	issn = {1748-9326},
	url = {https://dx.doi.org/10.1088/1748-9326/ab38d3},
	doi = {10.1088/1748-9326/ab38d3},
	language = {en},
	number = {9},
	urldate = {2025-04-08},
	journal = {Environmental Research Letters},
	author = {van der Wiel, Karin and Bloomfield, Hannah C and Lee, Robert W and Stoop, Laurens P and Blackport, Russell and Screen, James A and Selten, Frank M},
	month = sep,
	year = {2019},
	note = {Publisher: IOP Publishing},
	pages = {094010},
}

@article{gotske_designing_2024,
	title = {Designing a sector-coupled {European} energy system robust to 60 years of historical weather data},
	volume = {15},
	copyright = {2024 The Author(s)},
	issn = {2041-1723},
	url = {https://www.nature.com/articles/s41467-024-54853-3},
	doi = {10.1038/s41467-024-54853-3},
	language = {en},
	number = {1},
	urldate = {2025-04-08},
	journal = {Nature Communications},
	author = {Gøtske, Ebbe Kyhl and Andresen, Gorm Bruun and Neumann, Fabian and Victoria, Marta},
	month = dec,
	year = {2024},
	note = {Publisher: Nature Publishing Group},
	pages = {10680},
}

@article{pfenninger_dealing_2017,
	title = {Dealing with multiple decades of hourly wind and {PV} time series in energy models: {A} comparison of methods to reduce time resolution and the planning implications of inter-annual variability},
	volume = {197},
	issn = {0306-2619},
	shorttitle = {Dealing with multiple decades of hourly wind and {PV} time series in energy models},
	url = {https://www.sciencedirect.com/science/article/pii/S0306261917302775},
	doi = {10.1016/j.apenergy.2017.03.051},
	urldate = {2025-04-08},
	journal = {Applied Energy},
	author = {Pfenninger, Stefan},
	month = jul,
	year = {2017},
	pages = {1--13},
}

@article{grochowicz_using_2024,
	title = {Using power system modelling outputs to identify weather-induced extreme events in highly renewable systems},
	volume = {19},
	issn = {1748-9326},
	url = {https://dx.doi.org/10.1088/1748-9326/ad374a},
	doi = {10.1088/1748-9326/ad374a},
	language = {en},
	number = {5},
	urldate = {2025-04-08},
	journal = {Environmental Research Letters},
	author = {Grochowicz, Aleksander and van Greevenbroek, Koen and Bloomfield, Hannah C},
	month = apr,
	year = {2024},
	note = {Publisher: IOP Publishing},
	pages = {054038},
}

@article{levin_energy_2023,
	title = {Energy storage solutions to decarbonize electricity through enhanced capacity expansion modelling},
	volume = {8},
	copyright = {2023 Springer Nature Limited},
	issn = {2058-7546},
	url = {https://www.nature.com/articles/s41560-023-01340-6},
	doi = {10.1038/s41560-023-01340-6},
	language = {en},
	number = {11},
	urldate = {2025-04-08},
	journal = {Nature Energy},
	author = {Levin, Todd and Bistline, John and Sioshansi, Ramteen and Cole, Wesley J. and Kwon, Jonghwan and Burger, Scott P. and Crabtree, George W. and Jenkins, Jesse D. and O’Neil, Rebecca and Korpås, Magnus and Wogrin, Sonja and Hobbs, Benjamin F. and Rosner, Robert and Srinivasan, Venkat and Botterud, Audun},
	month = nov,
	year = {2023},
	note = {Publisher: Nature Publishing Group},
	pages = {1199--1208},
}

@article{luderer_impact_2022,
	title = {Impact of declining renewable energy costs on electrification in low-emission scenarios},
	volume = {7},
	copyright = {2021 The Author(s), under exclusive licence to Springer Nature Limited},
	issn = {2058-7546},
	url = {https://www.nature.com/articles/s41560-021-00937-z},
	doi = {10.1038/s41560-021-00937-z},
	language = {en},
	number = {1},
	urldate = {2025-04-08},
	journal = {Nature Energy},
	author = {Luderer, Gunnar and Madeddu, Silvia and Merfort, Leon and Ueckerdt, Falko and Pehl, Michaja and Pietzcker, Robert and Rottoli, Marianna and Schreyer, Felix and Bauer, Nico and Baumstark, Lavinia and Bertram, Christoph and Dirnaichner, Alois and Humpenöder, Florian and Levesque, Antoine and Popp, Alexander and Rodrigues, Renato and Strefler, Jessica and Kriegler, Elmar},
	month = jan,
	year = {2022},
	note = {Publisher: Nature Publishing Group},
	pages = {32--42},
}

@article{breyer_history_2022,
	title = {On the {History} and {Future} of 100\% {Renewable} {Energy} {Systems} {Research}},
	volume = {10},
	issn = {2169-3536},
	url = {https://ieeexplore.ieee.org/document/9837910},
	doi = {10.1109/ACCESS.2022.3193402},
	urldate = {2025-04-08},
	journal = {IEEE Access},
	author = {Breyer, Christian and Khalili, Siavash and Bogdanov, Dmitrii and Ram, Manish and Oyewo, Ayobami Solomon and Aghahosseini, Arman and Gulagi, Ashish and Solomon, A. A. and Keiner, Dominik and Lopez, Gabriel and Østergaard, Poul Alberg and Lund, Henrik and Mathiesen, Brian V. and Jacobson, Mark Z. and Victoria, Marta and Teske, Sven and Pregger, Thomas and Fthenakis, Vasilis and Raugei, Marco and Holttinen, Hannele and Bardi, Ugo and Hoekstra, Auke and Sovacool, Benjamin K.},
	year = {2022},
	pages = {78176--78218},
}

@article{hansen_status_2019,
	title = {Status and perspectives on 100\% renewable energy systems},
	volume = {175},
	issn = {0360-5442},
	url = {https://www.sciencedirect.com/science/article/pii/S0360544219304967},
	doi = {10.1016/j.energy.2019.03.092},
	urldate = {2025-04-08},
	journal = {Energy},
	author = {Hansen, Kenneth and Breyer, Christian and Lund, Henrik},
	month = may,
	year = {2019},
	pages = {471--480},
}

@article{brown_response_2018,
	title = {Response to ‘{Burden} of proof: {A} comprehensive review of the feasibility of 100\% renewable-electricity systems’},
	volume = {92},
	issn = {1364-0321},
	shorttitle = {Response to ‘{Burden} of proof},
	url = {https://www.sciencedirect.com/science/article/pii/S1364032118303307},
	doi = {10.1016/j.rser.2018.04.113},
	urldate = {2025-04-08},
	journal = {Renewable and Sustainable Energy Reviews},
	author = {Brown, T. W. and Bischof-Niemz, T. and Blok, K. and Breyer, C. and Lund, H. and Mathiesen, B. V.},
	month = sep,
	year = {2018},
	pages = {834--847},
}

@misc{international_energy_agency_net_2024,
	title = {Net {Zero} {Roadmap}: {A} {Global} {Pathway} to {Keep} the 1.5 °{C} {Goal} in {Reach} - 2023 {Update}},
	language = {en},
	author = {International Energy Agency},
	year = {2024},
}

@incollection{intergovernmental_panel_on_climate_change_ipcc_technical_2023,
	edition = {1},
	title = {Technical {Summary}},
	copyright = {https://www.cambridge.org/core/terms},
	isbn = {978-1-009-15792-6},
	url = {https://www.cambridge.org/core/product/identifier/9781009157926%23pre3/type/book_part},
	language = {en},
	urldate = {2025-04-08},
	booktitle = {Climate {Change} 2022 - {Mitigation} of {Climate} {Change}},
	publisher = {Cambridge University Press},
	editor = {{Intergovernmental Panel On Climate Change (Ipcc)}},
	month = aug,
	year = {2023},
	doi = {10.1017/9781009157926.002},
	pages = {51--148},
}

@article{pan_long-term_2022,
	title = {Long-term thermal performance analysis of a large-scale water pit thermal energy storage},
	volume = {52},
	issn = {2352-152X},
	url = {https://www.sciencedirect.com/science/article/pii/S2352152X22010064},
	doi = {10.1016/j.est.2022.105001},
	urldate = {2024-10-29},
	journal = {Journal of Energy Storage},
	author = {Pan, Xinyu and Xiang, Yutong and Gao, Meng and Fan, Jianhua and Furbo, Simon and Wang, Dengjia and Xu, Chao},
	month = aug,
	year = {2022},
	pages = {105001},
}

@article{sifnaios_impact_2023,
	title = {The impact of large-scale thermal energy storage in the energy system},
	volume = {349},
	issn = {0306-2619},
	url = {https://www.sciencedirect.com/science/article/pii/S0306261923010279},
	doi = {10.1016/j.apenergy.2023.121663},
	urldate = {2024-10-29},
	journal = {Applied Energy},
	author = {Sifnaios, Ioannis and Sneum, Daniel Møller and Jensen, Adam R. and Fan, Jianhua and Bramstoft, Rasmus},
	month = nov,
	year = {2023},
	pages = {121663},
}

@article{jacob_future_2023,
	title = {The future role of thermal energy storage in 100\% renewable electricity systems},
	volume = {4},
	issn = {2667-095X},
	url = {http://dx.doi.org/10.1016/j.rset.2023.100059},
	doi = {10.1016/j.rset.2023.100059},
	language = {en},
	journal = {Renewable and Sustainable Energy Transition},
	author = {Jacob, Rhys and Hoffmann, Maximilian and Weinand, Jann Michael and Linßen, Jochen and Stolten, Detlef and Müller, Michael},
	month = aug,
	year = {2023},
	pages = {100059},
}

@article{zeyen_mitigating_2021,
	title = {Mitigating heat demand peaks in buildings in a highly renewable {Europe} an energy system},
	volume = {231},
	issn = {0360-5442},
	url = {http://dx.doi.org/10.1016/j.energy.2021.120784},
	doi = {10.1016/j.energy.2021.120784},
	language = {en},
	journal = {Energy},
	author = {Zeyen, Elisabeth and Hagenmeyer, Veit and Brown, Tom},
	month = sep,
	year = {2021},
	pages = {120784},
}
